\documentclass[a4paper,fleqn,usenatbib]{mnras}

\usepackage{newtxtext,newtxmath}

\usepackage[T1]{fontenc}
\usepackage{ae,aecompl}

\usepackage{graphicx}	
\usepackage{amsmath}	
\usepackage{amssymb}	

\usepackage{verbatim}
\usepackage[dvipsnames]{xcolor}
\usepackage{units}
\usepackage{textcomp}
\usepackage{colortbl}

\usepackage{tikz}
\usepackage{pgfplots}
\usepackage{filecontents}

\usepackage{hyperref}
\usepackage{cleveref}

\title[Intrinsic polarisation of elongated porous dust grains]{Intrinsic polarisation of elongated porous dust grains}

\author[F. Kirchschlager et al.]{Florian Kirchschlager$^{1}$\thanks{E-mail: f.kirchschlager@ucl.ac.uk}, Gesa H.-M. Bertrang$^{2}$ and Mario Flock$^{2}$\\
 $^{1}$Department of Physics and Astronomy, University College London, Gower Street, London WC1E 6BT, United Kingdom\\
 $^{2}$Max Planck Institute for Astronomy, K\"onigstuhl 17, 69117 Heidelberg, Germany
 }

\date{Accepted 2019 June 25. Received 2019 June 17; in original form 2019 June 06}

\pubyear{2019}

\begin{document}
\label{firstpage}
\pagerange{\pageref{firstpage}--\pageref{lastpage}}
\maketitle

 \begin{abstract}
ALMA observations revealed recently polarised radiation of several protoplanetary disks in the (sub-)millimetre wavelength range. Besides self-scattering of large particles, thermal emission by elongated grains is a potential source for the detected polarisation signal. We calculate the wavelength dependent absorption and intrinsic polarisation of spheroidally shaped, micrometre and sub-millimetre sized dust grains using the discrete dipole approximation. In particular, we analyse the impact of dust grain porosity which appears to be present in disks  when small grains coagulate to form larger aggregates. For the first time our results show that (a) the intrinsic polarisation decreases for increasing grain porosity and (b) the polarisation orientation flips by 90 degree for certain ratios of wavelength to grain size. We present a new method to constrain grain porosity and the grain size in protoplanetary disks using multi-wavelength polarisation observations in the far-infrared to millimetre wavelengths. Finally, we find that moderate grain porosities ($\mathcal{P}\lesssim0.7$) potentially explain the observed polarisation fraction in the system HD~142527 while highly porous grains ($\mathcal{P}>0.7$) fail unless the grain's axis ratio is extraordinarily large.
 \end{abstract}

 


\section{Introduction}
\label{101}
Dust grains in protoplanetary disks are responsible for the absorption and scattering of both stellar and thermal radiation (in the optically thick regime). Besides the geometrical characteristics of these disks, dust grain properties such as their size, composition and structure are the decisive elements for the optical appearance of the disks. Therefore, the disk's emitted and scattered light also carries information about the disk's building blocks -- in total but also in polarised radiation.

\cite{Tamura1999} first detected polarised emission from T~Tauri disks using SCUBA/JCMT. This polarisation has been interpreted as thermal emission by magnetically aligned dust grains. Following detections of polarised signals from protoplanetary disks using SMA, CARMA, and VLA (\citealt{Rao2014,Stephens2014,Seguracox2015}), polarisation observations at $\sim$ millimetre wavelengths have made a further leap forward since high-resolved imaging of ALMA came available (\citealt{Kataoka2016, Kataoka2017, Cox2018, Girart2018, Harris2018, Hull2018, Lee2018, Ohashi2018, Sadavoy2018,Dent2019, Harrison2019, Takahashi2019}). The measured polarisation signal is strongly influenced by the resolution of the observation, projection along the line of sight, dust properties, and (if applicable) alignment mechanisms of aspherical grains, and the interpretation of the polarised observations requires sophisticated models (\citealt{Cho2007, Bertrang2017a, Bertrang2017b}). 

Accordingly, the origin of the detected polarisation signals is still highly debated. Besides dust scattering of anisotropic, thermal radiation of grains  of size $\sim\lambda/(2\,\pi)$ (\citealt{Kataoka2015}), dichroic extinction and emission of aligned grains is supposed to be accountable for the polarised radiation (\citealt{Cho2007,Bertrang2017b}). When light passes an elongated grain, the component of the electric vector parallel and perpendicular to the short axis of the grain is differently absorbed which causes a net polarisation, namely intrinsic (emission) polarisation. Considering the huge number of dust grains in a disk, their individual polarisation states would completely cancel out for disordered grains. However, grains can get aligned by several factors: the magnetic field (\citealt{Lazarian2007}), the radiation field (\citealt{Lazarian2007, Tazaki2017}), or mechanically (aerodynamically; \citealt{Gold1952}).

So far, all studies considering polarisation of elongated particles assumed solid dust grains 
(\citealt{Draine1996, Draine1997, Cho2007,Lazarian2007}). It is well known that dust grains grow by coagulation in protoplanetary disks, and as a result they might be porous. The occurrence of grain porosity is not only predicted by theoretical dust growth studies (e.g.~\citealt{Dominik1997, Ormel2007}) but also by laboratory experiments\break (e.g.~\citealt{BlumWurm2008, Kothe2013}) as well as observations of circumstellar disks  (e.g.~\citealt{Pinte2008, Milli2019}). Moreover, the \textit{Rosetta} orbiter detected fluffy grains in situ of the comet \mbox{67P/Churyumov-Gerasimenko} (\citealt{Fulle2015,Langevin2016,Mannel2016}). Due to dust settling of more compact grains, porous grains are expected to occur cumulatively close to the disk surface, enhancing their contribution to the disk's optical appearance. Further, porosity has a huge impact on the grain's optical properties (e.g.~\citealt{Kirchschlager2013,Kirchschlager2014, Tazaki2018, Ysard2018}). Therefore, the study of elongated porous grains is promising to contribute to the comprehensive understanding of polarisation by thermal emission. 

In this work, we calculate the optical properties of spheroidally porous dust grains. In Section~\ref{201} we present the model for our dust grains and outline the calculation method. In Section~\ref{301} we discuss the wavelength dependent absorption and intrinsic polarisation of spheroidally porous grains, for single particles as well as for a grain size distribution. We discuss observed polarisation fractions and introduce a procedure to derive the grain porosity from intrinsic polarisation observations in Section~\ref{501} and give a summary in Section~\ref{401}.

\section{Dust model and methods}
\label{201}
\subsection{Elongated porous dust grains}
\begin{figure}
 \centering
   \includegraphics[trim=0cm 0cm 0cm 0cm, clip=true,page=1,width=1.0\linewidth]{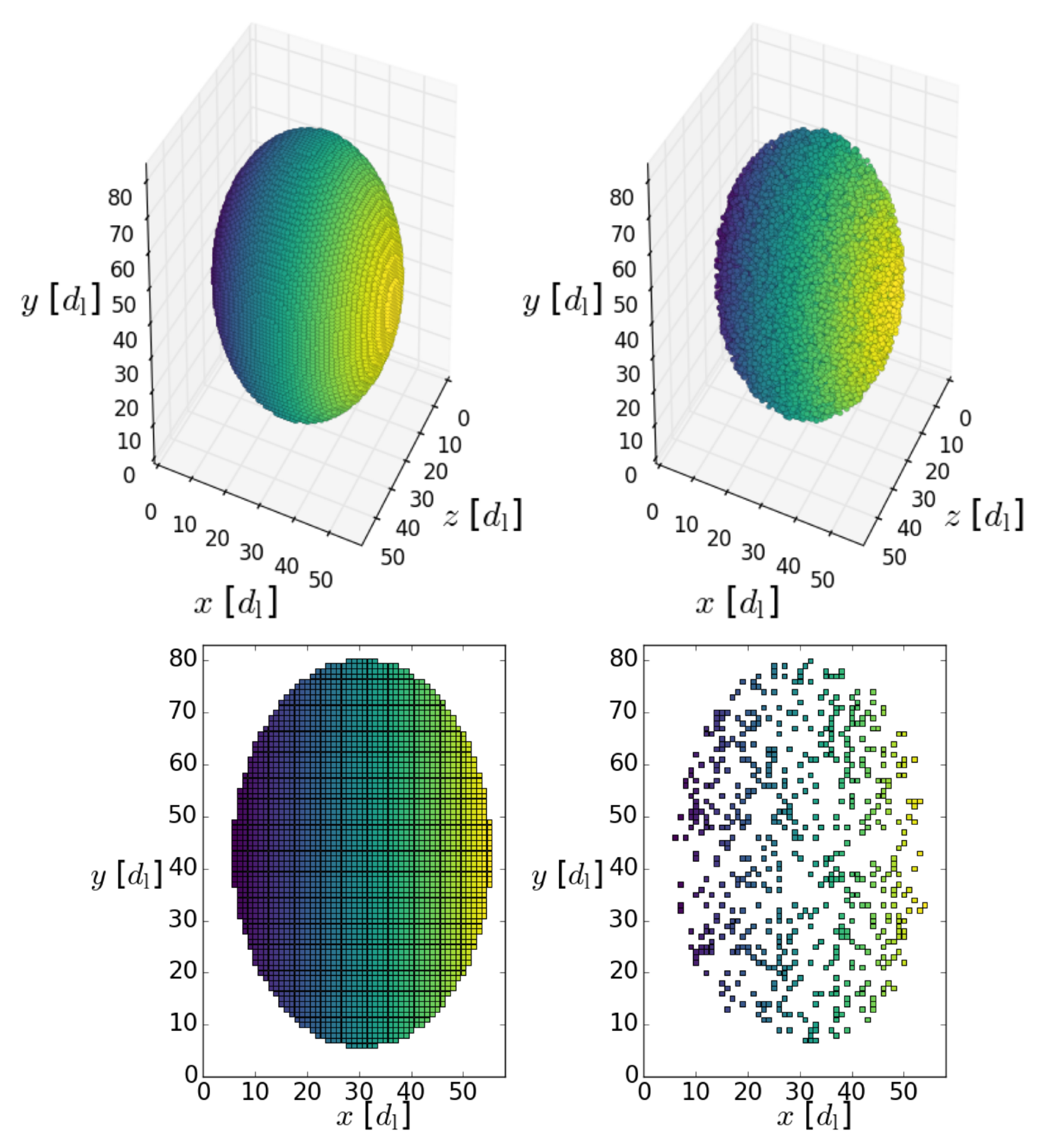}
\caption{Morphology of elongated dust grains (\mbox{1-1.5-1}). \textit{Top:} A compact grain composed of 96912 cubic subvolumes (\textsl{left}; \mbox{$\mathcal{P}=0.0$}) and a porous grain of 19382 cubic subvolumes (\textsl{right}; \mbox{$\mathcal{P}=0.8$}). \textit{Bottom:} Single-layer of cubic dipoles through the centre of the compact and porous elongated grain, respectively. The colour scale is only for illustration purposes.}
\label{fig_morph_DDA} 
\end{figure}

In our study the elongated porous dust grains have a spheroidally basic shape from which material is removed so that voids are formed. This model is comparable to that from \cite{Kirchschlager2013,Kirchschlager2014} where a spherically basic form is used. Here, we consider 6 spheroidally basic shapes with short semi-axis $b$ and long semi-axis $c$: 3 prolate and 3 oblate shapes with $\nicefrac{c}{b}\in\{1.1,1.3,1.5\}$, respectively. A prolate particle shape will be referred to as `$1-\nicefrac{c}{b}-1$'  and an oblate shape as `$\nicefrac{c}{b}-1-\nicefrac{c}{b}$'. The rotational axis of all spheroids is parallel to the $y$-axis  in a Cartesian coordinate system ($x-y-z$; Fig.~\ref{fig_morph_DDA}). 

Porosity is the ratio of the vacuum volume to the total volume of the spheroidally grain, 
\begin{align}
 \mathcal{P}=V_\text{vacuum}/V_\text{total} = 1-V_\text{solid}/V_\text{total} . 
\end{align}
Here, $V_\text{total}$ is the volume of the encasing spheroid,  \mbox{$V_\text{total}=4/3\pi b^2 c$} for prolate and $V_\text{total}=4/3\pi bc^2$ for oblate grains. The effective radius is the radius of a volume-equivalent solid sphere, $a_\text{eff}=\left(b^2c(1-\mathcal{P})\right)^{1/3}$ for prolate and $a_\text{eff}=\left(bc^2(1-\mathcal{P})\right)^{1/3}$ for oblate grains. We note that two grains composed of the same dust material and with the same effective radius $a_\text{eff}$ have the same mass.

Our model enables the description of elongated porous dust grains by a low number of parameters ($a_\text{eff}, \nicefrac{c}{b}, \mathcal{P}$) while the definition of the porosity is still accurate. Contrary, the porosity definition in many other studies suffers by the fact that the total volume is that of a sphere encasing the dust particle, even for highly aspherical grains such as BCCA aggregates (e.g.~\citealt{Kataoka2014, Tazaki2018}). When the encompassing body is a sphere, the porosity is artificially high; for example a solid but prolate dust grain with $\nicefrac{c}{b}=10$ (compact needle) would have a porosity of \mbox{$\unit[\sim99]{\%}$}. In contrast, our definition of porosity gives the correct value of $\mathcal{P}=0$ and an overestimation of the porosity is excluded.

\subsection{Calculation method}
\label{sec_cal_met}
To calculate the optical properties of irregular shaped particles we use the code \textsc{DDSCAT}\footnote{\href{https://www.astro.princeton.edu/~draine/DDSCAT.7.3.html}{https://www.astro.princeton.edu/~draine/DDSCAT.7.3.html}} (version 7.3; \citealt{Draine1994,Draine2010}) which is based on the theory of discrete dipole approximation (DDA; \citealt{PurcellPenny1973}). The three-dimensional particle shape is replaced by a corresponding spatial distribution of $N$ discrete dipoles on a cubic grid and the optical properties are then calculated for this dipole distribution. The \textsc{DDSCAT} is well tested and is applicable for most particle shapes and structures. However, the applicability is constrained by an upper limit of the ratio of grain size $a_\text{eff}$ to wavelength $\lambda$ (see e.g.~\citealt{DraineGood, Draine1994, Draine2000}). For porous dust grains, this condition is given by 
(\citealt{Kirchschlager2013})
\begin{align}
 \frac{a_\text{eff}}{\lambda} \lesssim 0.1\frac{N^{1/3}}{|n(\lambda)|},
\end{align}
where $n(\lambda)$ is the wavelength dependent complex refractive index. The lowest number of dipoles used in this study is $N= 19382$, and with $|n(\lambda)|\sim 3$ (for astronomical silicate at $\lambda=\unit[100]{\mu m}$)\footnote{For carbon and ice is $|n(\lambda)|\sim 1.8$  at $\lambda=\unit[100]{\mu m}$, and it follows $\lambda\gtrsim 0.7 a_\text{eff}$.}  follows: $\lambda\gtrsim a_\text{eff}$. As the ratio of the effective radius $a_\text{eff}$ to wavelength $\lambda$ increases, so does the necessary computation time. For the 1 and 10 micrometre grains the optical properties can be calculated in less than $2.5$ core hours using a $\unit[2.70]{GHz}$ central processing unit (CPU), whereas the calculations for a 100 micrometre grain took approximately $7.5$ days.

\subsection{Porous structure}
The method to generate the porous dust grains is described in detail in \cite{Kirchschlager2013} and briefly summarised here: $N'$ dipoles are arranged on a cubic lattice with lattice constant $d_\text{l}$, forming the shape of a compact spheroid with semi-axis $b$ and $c$ and a total volume of \mbox{$N'\left(d_\text{l}/2\right)^3 = 4/3\pi a_\text{eff}^3\left(1-\mathcal{P}\right)^{-1}$.} In order to create a porous grain of porosity $\mathcal{P}$, $\mathcal{P}N'$ dipoles are randomly removed. Then the remaining dipole arrangement of $N=\left(1-\mathcal{P}\right)N'$ dipoles forms the elongated porous dust grain (Fig.~\ref{fig_morph_DDA}). Each dipole represents a cubic subvolume $d_\text{l}^3$ of the dust material, while a removed dipole represents a cubic void of the same size. The large number of dipoles ($\sim10^5-10^6$) ensures a statistical distribution of the voids within the grains and a good approximation of the spheroidally basic shape. 

\begin{figure}
 \centering
 \fbox{\includegraphics[trim=0cm 2cm 0cm 2.5cm,  clip=true,page=1,width=1.0\linewidth]{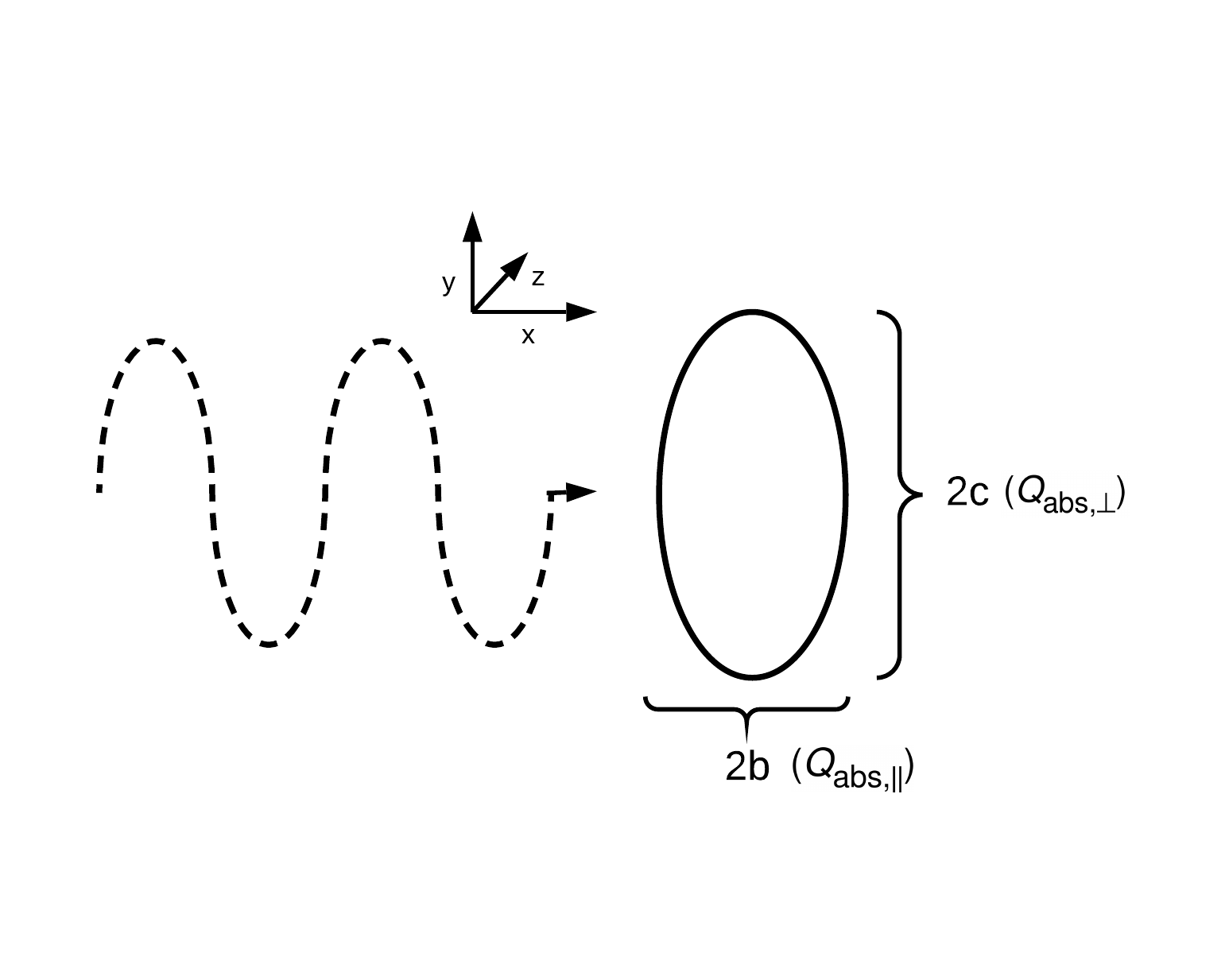}}
\caption{Sketch of the irradiation of an elongated particle.}
\label{fig_sketch_elong} 
\end{figure}

Using this method, the dipole configurations of six particle shapes and nine porosities ($\mathcal{P}=0.0-0.8$; in steps of $0.1$) are generated. We calculate the optical properties for each of the 54 ($=6\times 9$) dipole configurations for the grain sizes $a_\text{eff}=1,10$ and $\unit[100]{\mu m}$ in the wavelength range $\lambda\in\left[\unit[100]{\mu m},\unit[2000]{\mu m}\right]$ for the case that the radiation is propagating parallel to the $x$-direction (Fig.~\ref{fig_sketch_elong}). We consider grains of three different dust materials: astronomical silicate (\citealt{Draine2003a,Draine2003b}), carbon (\citealt{Jaeger1998}), and ice (\citealt{Warren1984, Reinert2015}). We have seen in Section~\ref{sec_cal_met} that the chosen number of dipoles and dust materials is sufficient to calculate the optical properties for dust grains up to $a_\text{eff}=\unit[100]{\mu m}$ for the shortest wavelengths $\lambda=\unit[100]{\mu m}$.


\section{Results}
\label{301}
\subsection{Polarised emission of single dust grains}
\begin{figure}
 \centering
 \includegraphics[trim=2.42cm 3.81cm 2cm 3.42cm,  clip=true,page=5,width=1.0\linewidth]{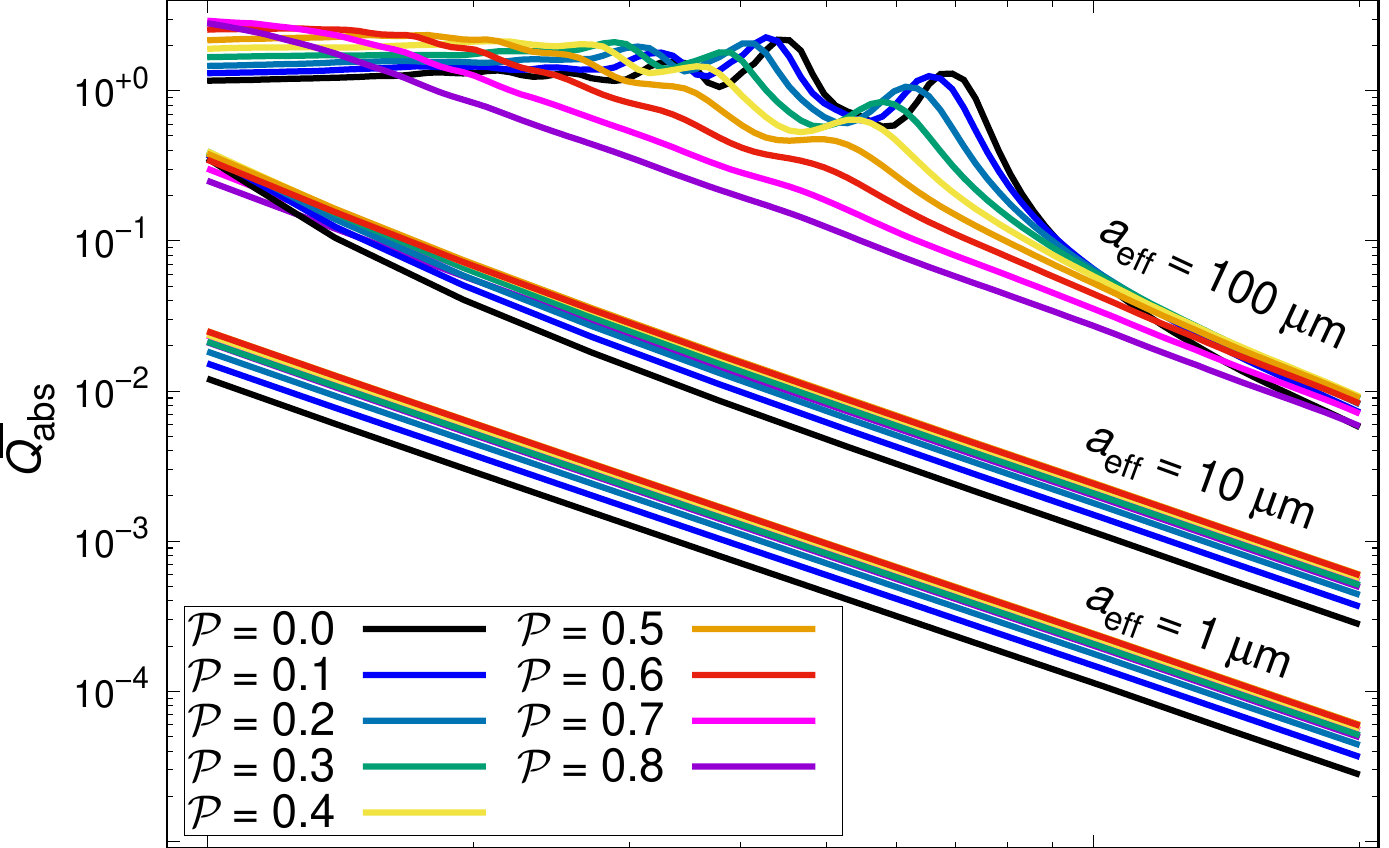}\\[-0.04cm]
 \includegraphics[trim=2.42cm 2.0cm 2cm 3.4cm,  clip=true,page=4,width=1.0\linewidth]{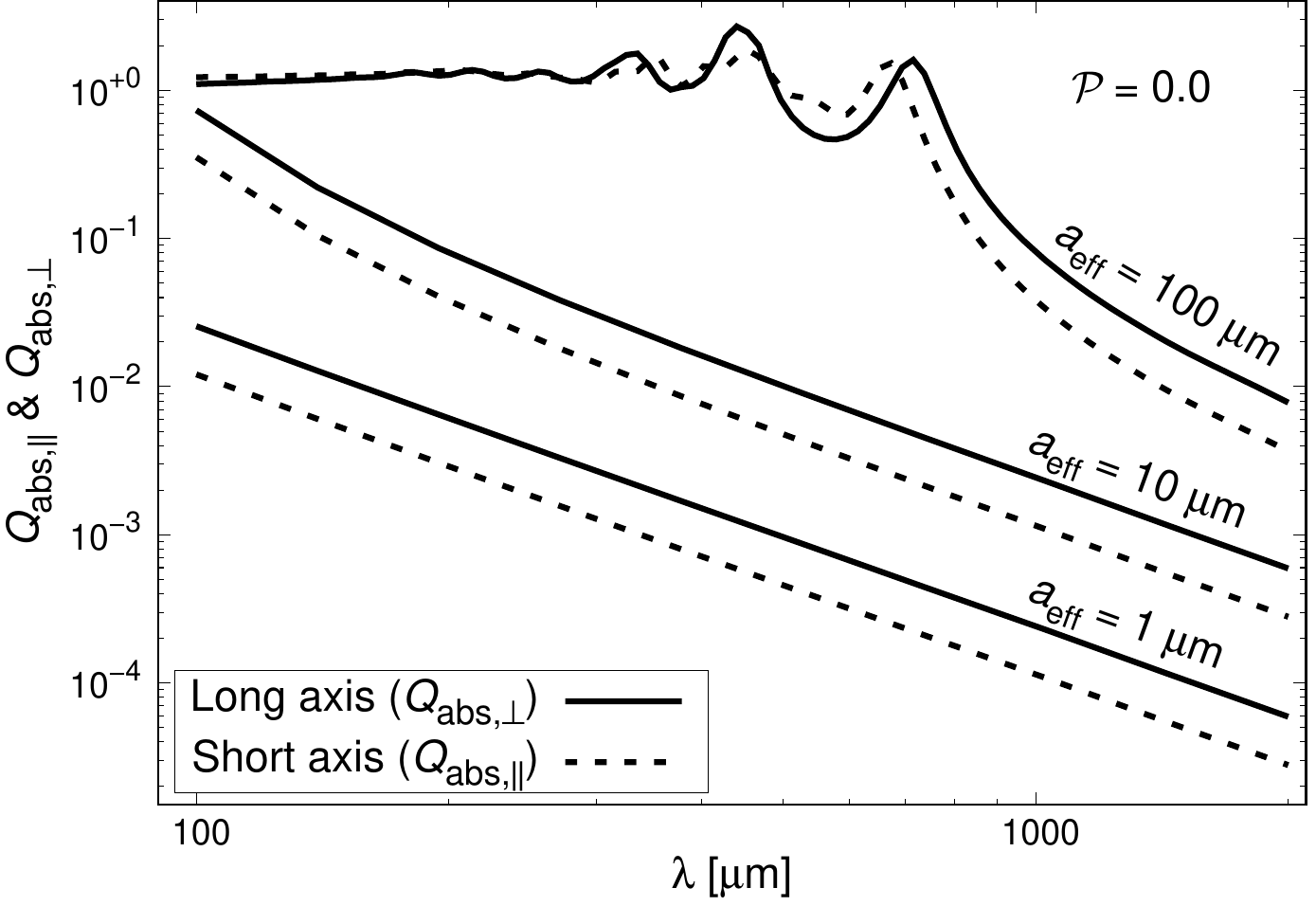}
\caption{Absorption efficiency as a function of wavelength $\lambda$. The dust grain is prolate (1-1.5-1), composed of astronomical silicate and has an effective dust radius of $a_\text{eff}=\unit[1]{\mu m}$, $\unit[10]{\mu m}$ and $\unit[100]{\mu m}$, respectively. \textit{Top:} Averaged absorption $\overline{Q}_\text{abs}$ as a function of porosity~$\mathcal{P}$. \textit{Bottom:} Absorption of the long and short axis for compact spheroids ($\mathcal{P}=0.0$).}
\label{fig_qabs} 
\end{figure}
When the rotational axis of the elongated grain is parallel to the $y$-direction and the photons travel parallel to $x$-direction, photons ``see'' an ellipsoidally geometric cross section of the grain defined by the short and long semi-axis, $b$ and $c$, respectively. The electric field parallel to the short axis sees a smaller cross section than that perpendicular to the short axis. Consequently, the efficiency factors of absorption $Q_{\text{abs},\parallel}$ and $Q_{\text{abs},\bot}$ can be differentiated. In Fig.~\ref{fig_qabs}, $Q_{\text{abs},\parallel}$ and $Q_{\text{abs},\bot}$ are shown for compact silicate spheroids of shape 1-1.5-1 as well as the averaged absorption $\overline{Q}_{\text{abs}} = 0.5\left(Q_{\text{abs},\parallel} + Q_{\text{abs},\bot}\right)$ for different porosities. One can see that the absorption perpendicular to the short axis is larger than that parallel to the short axis for wavelengths $\lambda>2\pi a_\text{eff}$. Furthermore, the absorption efficiency increases with increasing porosity for wavelengths $\lambda\lesssim2a_\text{eff}$ or $\lambda\gtrsim 10a_\text{eff}$, while the order is reversed at intermediate wavelengths. We note that we compare grains of different porosity but with the same effective radius $a_\text{eff}$. Therefore, the grains have the same mass but a different spatial extension ($V_\text{total}$). However, we show in Appendix~\ref{sec_same_shell} that the dependence of $\overline{Q}_{\text{abs}}$ on the porosity is still present if grains with same spatial extension but different mass are compared.

Since the absorption is different for the long and the short axis, the radiation gets polarised. Moreover, the efficiency factor of emission is equal to that of absorption ($Q_\text{emi}=Q_\text{abs}$), and radiation emitted by elongated dust grains is polarised in the same way. The degree of intrinsic polarisation by emission from elongated dust grains is then (\citealt{Cho2007})
\begin{align}
  P_\text{emi} = \frac{Q_{\text{abs},\bot}-Q_{\text{abs},\parallel}}{Q_{\text{abs},\bot} + Q_{\text{abs},\parallel}}.
\end{align}
 
\begin{figure*}
 \centering
 \includegraphics[trim=0cm 0cm 0cm 0cm, clip=true,width=1.0\linewidth]{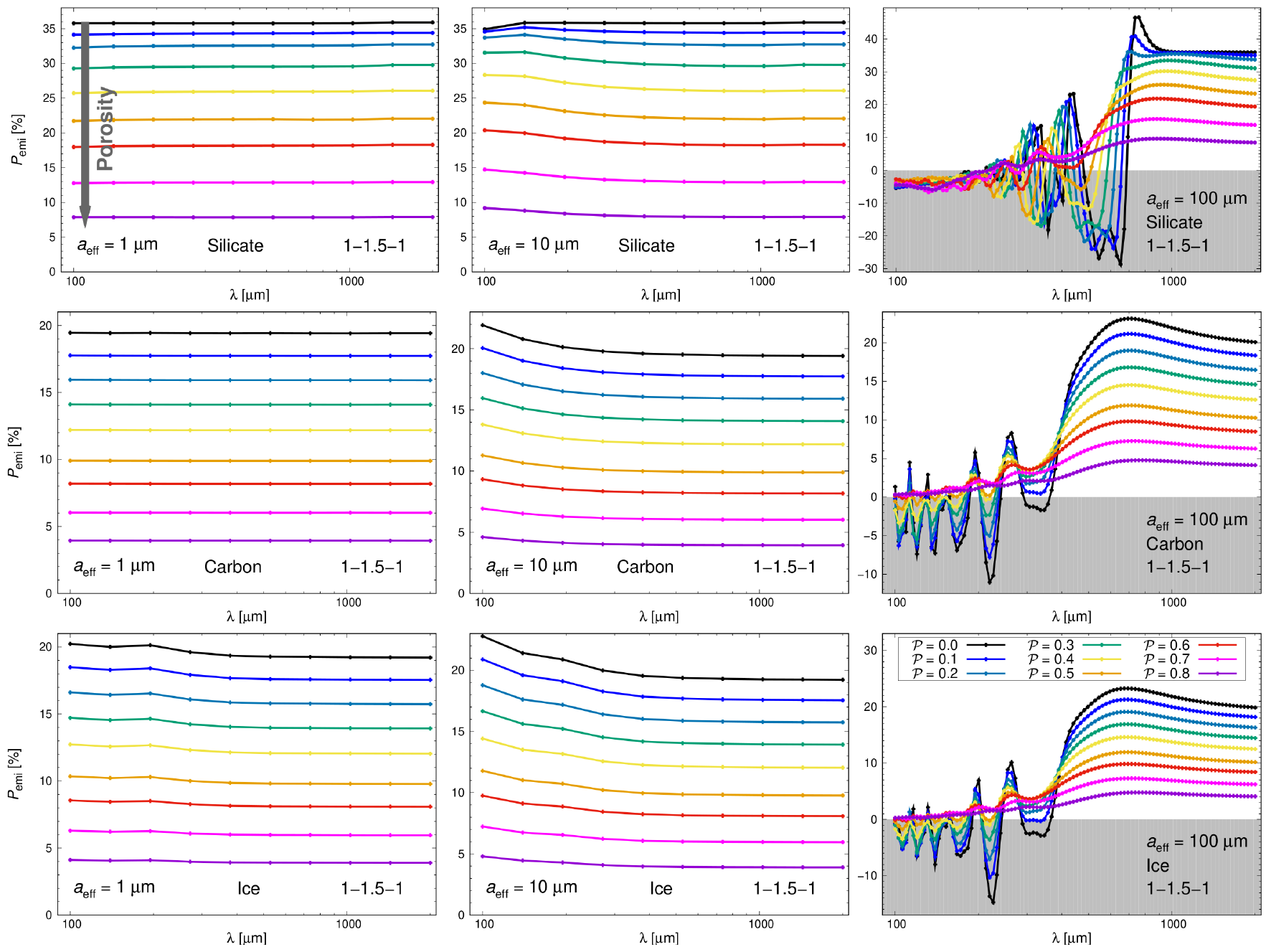} 
\caption{Intrinsic polarisation $P_\text{emi}$ as a function of wavelength $\lambda$, porosity~$\mathcal{P}$, and effective grain radii $a_\text{eff}=\unit[1]{\mu m}$, $\unit[10]{\mu m}$ and $\unit[100]{\mu m}$ for prolate (1-1.5-1) dust grains. The dust material is astronomical silicate (\textit{top} row), carbon (\textit{middle}), and ice (\textit{bottom}). The grey coloured regions emphasize negative polarisations which cause a 90 degree-flip of the polarisation direction.}
\label{fig_pol_emi} 
 \end{figure*}

\begin{figure*}
  \centering
  \includegraphics[trim=2.3cm 2.2cm 2cm 2.3cm, clip=true,page=1,height=0.251\linewidth]{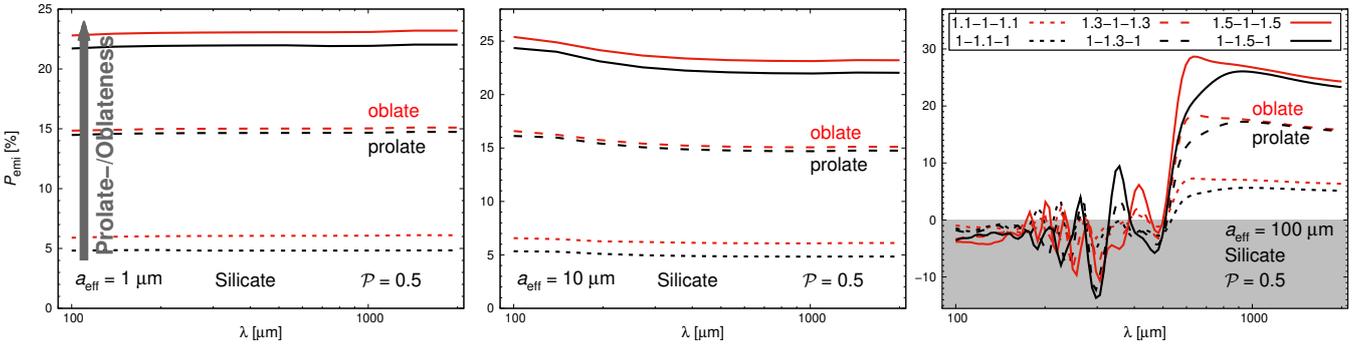}\hspace*{-0.01cm}
  \includegraphics[trim=3.8cm 2.2cm 2cm 2.3cm, clip=true,page=2,height=0.251\linewidth]{Polarisation_obl_vs_prol.pdf}\hspace*{-0.01cm}
  \includegraphics[trim=3.8cm 2.2cm 2cm 2.3cm, clip=true,page=3,height=0.251\linewidth]{Polarisation_obl_vs_prol.pdf}
 \caption{Intrinsic polarisation $P_\text{emi}$ as a function of prolate- (black) or oblateness (red) for elongated silicate grains with axis ratio  $\nicefrac{c}{b}=1.1, 1.3$, or $1.5$ (dotted, dashed, solid lines). The porosity is fixed to $\mathcal{P}=0.5$ and the material is astronomical silicate.}
\label{fig_pol_emi_obl_vs_prol} 
\end{figure*}

The degree of intrinsic polarisation of prolate silicate grains (1-1.5-1)  is shown in Fig.~\ref{fig_pol_emi} (\textit{top} row) as a function of wavelength $\lambda$, porosity $\mathcal{P}$, and effective grain size $a_\text{eff}$. For the small and medium sized grains ($a_\text{eff}=\unit[1]{\mu m}$ and $\unit[10]{\mu m}$), we can see the following:

\begin{description}
 \item [\textbf{Porosity}] The degree of intrinsic polarisation $P_\text{emi}$ is decreasing with increasing porosity $\mathcal{P}$ for all particle shapes. This is surprising, as one would expect intuitively an increase of polarisation with increasing porosity. Moreover, the polarisation by scattering of porous grains increases with porosity in the optical wavelength range (\citealt{Kirchschlager2014}).
 \item [\textbf{Wavelength}] In the far-infrared to millimetre wavelength range, the intrinsic polarisation $P_\text{emi}$ is nearly independent of the wavelength $\lambda$. Only for $a_\text{eff}=\unit[10]{\mu m}$ and at wavelengths $\lambda\sim\unit[100]{\mu m}$, a weak wavelength-dependence is visible for the polarisation.
\item [\textbf{Axis ratio}] The intrinsic polarisation $P_\text{emi}$  increases with increasing aspherity of the particles (Fig.~\ref{fig_pol_emi_obl_vs_prol}), as the ratio of long to short semi-axis $\nicefrac{c}{b}$ determines the absorption cross sections $Q_{\text{abs},\bot}$ and $Q_{\text{abs},\parallel}$. The longer prolate grains or the flatter oblate grains, the larger the polarisation.
 \item [\textbf{Prolate/oblate}] The intrinsic polarisation $P_\text{emi}$ is slightly higher for oblate compared to prolate grains of the same axis ratio $\nicefrac{c}{b}$, however, the difference is for most wavelengths and grain sizes below $\unit[2]{\%}$ (Fig.~\ref{fig_pol_emi_obl_vs_prol}). Therefore, the ratio of long to short semi-axis $\nicefrac{c}{b}$ and the porosity $\mathcal{P}$ are the only parameters of the particle morphology that have significant influence on the intrinsic polarisation $P_\text{emi}$.
\end{description}

For the $a_\text{eff}=\unit[100]{\mu m}$ grains (\textit{right} columns in Figs.~\ref{fig_pol_emi}  and \ref{fig_pol_emi_obl_vs_prol}), the polarisation degree shows a strong dependence on the wavelength. Oscillations occur as a result of enhanced interferences due to multiple reflection of the electromagnetic wave within the grain at $\lambda\sim2\,\pi a_\text{eff}$ (\citealt{vandeHulst1981}). The oscillations enable even ``negative polarisation'' which causes a flip of the polarisation direction, similar to the polarisation reversal by scattering in the optical regime (e.g. \citealt{Kirchschlager2014}). While the polarisation vector is parallel to the long grain axis (for both prolate and oblate grains) for $\lambda\gtrsim2\pi a_\text{eff}$, the polarisation direction is perpendicular to the long axis for smaller wavelengths. The amplitudes of the oscillations are slightly stronger for oblate grains compared to prolate grains but occur at similar wavelengths. Furthermore, the amplitudes of the oscillations decrease with increasing porosity as the voids hamper multiple reflection. For $\lambda\lesssim2\,a_\text{eff}$ or  $\lambda\gtrsim10\,a_\text{eff}$, the wavelength dependence vanishes completely. Although the oscillations complicate the trends observed for the small and medium sized grains, the curves for the $a_\text{eff}=\unit[100]{\mu m}$ grains are also ordered in a way that the amount of intrinsic polarisation is decreasing with increasing porosity and increasing with the ratio of long to short axis $\nicefrac{c}{b}$, at least for wavelengths $\lambda\gtrsim3\,a_\text{eff}$.

Regarding the grain size, three cases can be differentiated: for $\lambda\gtrsim10\,a_\text{eff}$ (Rayleigh regime), the effective radius $a_\text{eff}$ has no impact on the degree of polarisation and $P_\text{emi}$ is constant.  For $2\,a_\text{eff}\lesssim\lambda\lesssim10\,a_\text{eff}$, interferences cause large oscillations in both absorption efficiency and intrinsic polarisation. For $\lambda\lesssim2\,a_\text{eff}$ (geometric optics regime), photons do not recognise the elongated shape. The absorption parallel to the short axis approaches that of the long axis and the emission becomes unpolarised. 

Fig.~\ref{fig_pol_emi} shows also the degree of intrinsic polarisation $P_\text{emi}$ for grains composed of carbon (\citealt{Jaeger1998}) and ice (\citealt{Warren1984,Reinert2015}), respectively. The polarisations of these two materials are remarkably similar to each other. Compared to astronomical silicate, $P_\text{emi}$ is $\sim\unit[15]{\%}$ lower for carbon and ice, the oscillations occur at shorter wavelengths ($\lambda<4a_\text{eff}$) and at the same wavelengths for different porosities. Despite that, all trends observed for the intrinsic polarisation of astronomical silicate are also visible for carbon and ice.

To verify our results obtained with \textsc{DDSCAT} (based on DDA), we further performed simulations based on the \mbox{T-Matrix} method (TMM). We used the code \textsc{MSTM} (\citealt{Mackowski1996}) to calculate the absorption and hence the intrinsic polarisation for a configuration of multiple spheres arranged to approximate an elongated (porous) dust grain. The intrinsic polarisation obtained with \textsc{TMM} (Fig.~\ref{fig_compare_DDSCAT_MST}) shows the same trends as that obtained with \textsc{DDA}. For details we refer to Appendix~\ref{A101}.

\subsection{Polarised emission in a protoplanetary disk} 
Dust grains in protoplanetary disks occur in different sizes, and the resulting polarisation by emission will be composed of the contribution of each grain size. For an optically thin disk in which all dust grains are perfectly aligned, the degree of intrinsic polarisation is estimated by (c.f. \citealt{Cho2007})
\begin{align}
  P_\text{emi,disk}(\lambda) = \frac{\int_{a_\text{min}}^{a_\text{max}}\left[Q_{\text{abs},\bot}-Q_{\text{abs},\parallel}\right]a_\text{eff}^2n(a_\text{eff})\,\text{d}a_\text{eff}}{\int_{a_\text{min}}^{a_\text{max}}\left[ Q_{\text{abs},\bot} + Q_{\text{abs},\parallel}\right]a_\text{eff}^2n(a_\text{eff})\,\text{d}a_\text{eff}},
  \label{eq123}
\end{align}
where $\text{d}n(a_\text{eff})\sim a_\text{eff}^{-\gamma}\,\text{d}a_\text{eff}$ is the number of grains in the interval $[a_\text{eff},a_\text{eff}+\text{d}a_\text{eff}]$, $\gamma=3.5$ is the grain size exponent, and $a_\text{min}$ and $a_\text{max}$ are the minimum and maximum radii of the grain size distribution, respectively. We set $a_\text{min}=\unit[5]{nm}$ and $a_\text{max}=\unit[100]{\mu m}$ and make use of the fact that the degree of intrinsic polarisation is constant for $a_\text{eff}\le0.1\lambda$. For $a_\text{eff}>0.1\lambda$ we interpolate the results obtained for $a_\text{max}=\unit[10]{\mu m}$ and $\unit[100]{\mu m}$. Considering 1000 grain sizes logarithmically
equidistantly distributed in the range between $a_\text{min}$ and $a_\text{max}$, we calculate $P_\text{emi,disk}(\lambda)$ via equation~(\ref{eq123}). 

\begin{figure}
 \centering
  \includegraphics[trim=0cm 0cm 0.4cm 0.3cm,  clip=true,page=1,width=1.0\linewidth]{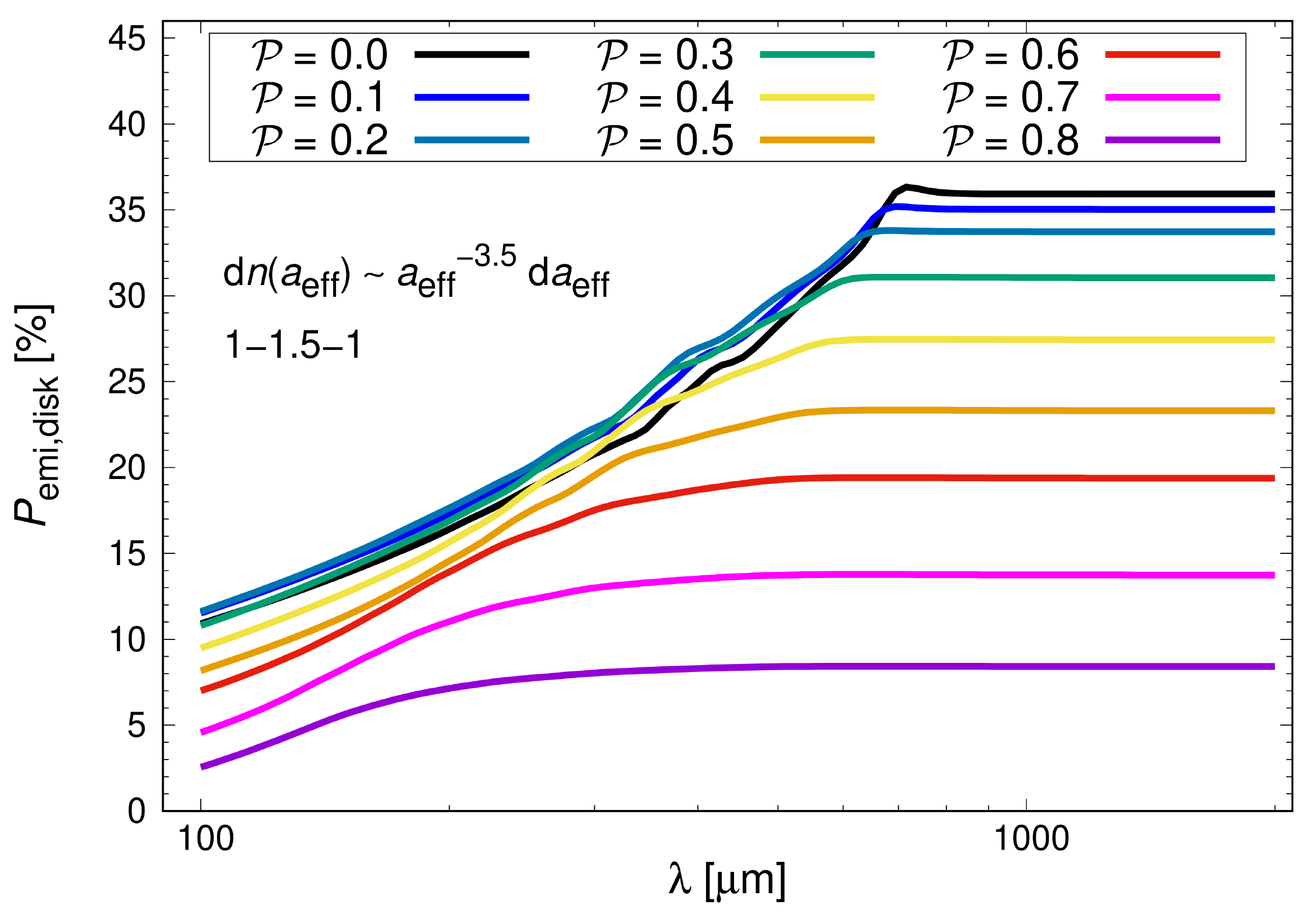}
\caption{Degree of polarisation by emission for an optically thin disk composed of 1-1.5-1 grains made of astronomical silicate, following a size distribution $\text{d}n(a_\text{eff})\sim a_\text{eff}^{-3.5}\text{d}a_\text{eff}$.}
\label{fig_pol_disk} 
\end{figure}

The resulting polarisation of the disk is presented in Fig.~\ref{fig_pol_disk} for grains of shape 1-1.5-1 made of astronomical silicate. For simplicity, we only present the result for this particle, but note that the figures are qualitatively similar for the other grain shapes and dust materials.\footnote{The intrinsic polarisation data for grain size distributions of other grain shapes and dust materials are online available as supplementary material.} While for single grains oscillations occur at wavelengths  $\lambda\in\left[\sim2\,a_\text{eff},10\,a_\text{eff}\right]$, these oscillations cancel out by the presence of different sizes. 

\begin{figure}
 \centering
 \includegraphics[trim=2.5cm 2.2cm 2.1cm 3.2cm,  clip=true,page=1,width=1.0\linewidth, page = 1]{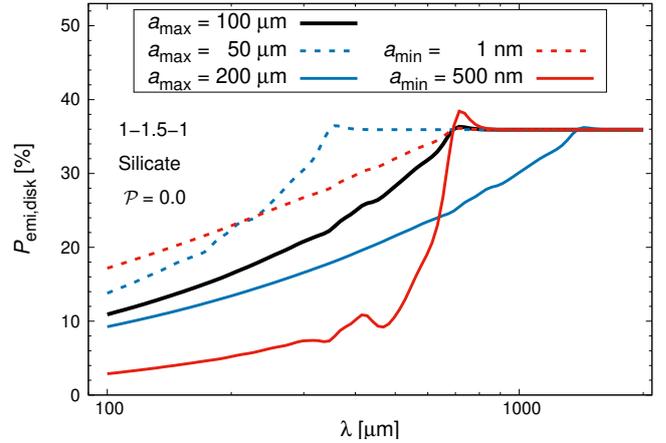}
\caption{Influence of maximum (blue) or minimum (red) grain size on the degree of intrinsic polarisation for silicate grains: the thick, black line represents the reference case  (\mbox{$a_\text{min}=\unit[5]{nm}$}; $a_\text{max}=\unit[100]{\mu m}$), dotted lines are the polarisation for a reduced minimum/maximum grain size, and the thin, solid lines are for an increased minimum/maximum grain size.}
\label{fig_pol_disk_change1} 
\end{figure}

\begin{figure}
 \centering
 \includegraphics[trim=2.5cm 2.2cm 2.1cm 3.2cm,  clip=true,page=1,width=1.0\linewidth, page = 2]{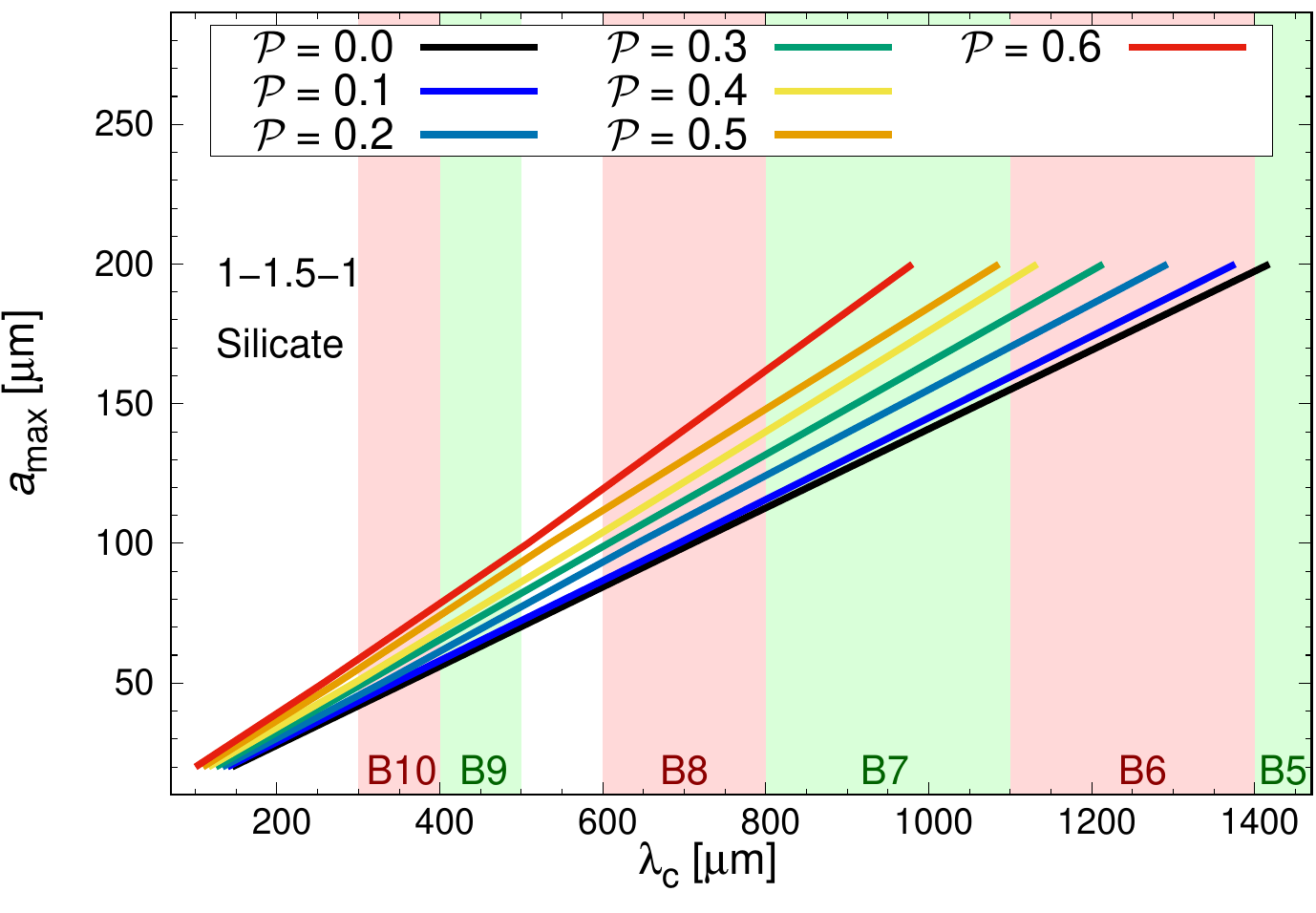}
\caption{Relation between the maximum grain radius $a_\text{max}$ of the grain size distribution  and the characteristic wavelength $\lambda_\text{c}$ at which the break from linearly increasing to constant polarisation $P_\text{emi,disk}$ occurs (knee). The coloured shaded regions represent the ALMA wavebands B5 to B10.}
\label{fig_pol_disk_change2} 
\end{figure}

Following the trends of the single grains, the intrinsic polarisation of the disk is decreasing with increasing porosity. We see that $P_\text{emi}$ is linearly increasing with $\log{(\lambda)}$ before the curve reaches a ``knee'' at a characteristic wavelength $\lambda_\text{c}$, and the polarisation is getting constant for wavelengths $\lambda>\lambda_\text{c}$. The reason for this break is the presence of grains which show oscillations for $\lambda<\lambda_\text{c}$, resulting in cancellation or reduction of the intrinsic polarisation, while all single grains show constant polarisation for $\lambda>\lambda_\text{c}$. As a consequence, when the maximum grain size $a_\text{max}$ is varied we find a linear shift of the knee and the relation $\lambda_\text{c} = 8\,a_\text{max}$ for compact silicate spheroids\footnote{For the standard MRN distribution (\citealt{Mathis1977}) with $a_\text{max}=\unit[0.25]{\mu m}$, we expect a constant intrinsic polarisation for all wavelengths in the interval $\lambda \in\left[\unit[100]{\mu m},\unit[2000]{\mu m}\right]$.} ($\mathcal{P}=0.0$; Fig.~\ref{fig_pol_disk_change1}). For larger porosities, the knee is less pronounced and shifted to shorter wavelengths (Fig.~\ref{fig_pol_disk_change2}).

We also varied the minimum grain size $a_\text{min}$ for a fixed maximum grain size $a_\text{max}$. While the values for the characteristic wavelength $\lambda_\text{c}$ are unaffected, the slope at small wavelengths is steeper for larger $a_\text{min}$ and shows oscillations, a consequence of the missing smoothing effect of the removed small particles. On the other hand, a reduction of the minimum grain size to lower values results in a flatter and smoother slope. Similar to that, we varied the grain size exponent $\gamma$ between 3 and 4 and find that values close to 3 (reduced number of small grains) cause larger oscillations, while \mbox{$\gamma  \sim 4$} gives a flatter and smoother slope too. In both cases, varying $a_\text{min}$ and $\gamma$ does not change the characteristic wavelength $\lambda_\text{c}$, which is only a function of the maximum grain size and porosity. Moreover, the characteristic wavelength is unaffected by the axis ratio $\nicefrac{c}{b}$ while it shifted to by a factor of 2 shorter wavelengths for carbon or ice composition.

We note that the presented polarisation values are derived under the assumption of perfectly aligned dust grains. However, grains in protoplanetary disks will be only partially aligned, and some parts of the disk will be optically thick, both reducing the amount of intrinsic polarisation (e.g.~\citealt{Cho2007, Bertrang2017a, Bertrang2017b}). 

\section{Implications for polarisation observations}
\label{501}
We have seen in Section~\ref{301} that the characteristic wavelength $\lambda_\text{c}$ only depends on the maximum grain size and the porosity but not on further grain size properties such as the axis ratio $\nicefrac{c}{b}$, the minimum grain size or the grain size exponent. As the maximum grain size can be derived from other observation techniques, e.g. from the (sub-)mm slope of the spectral energy distribution of the total intensity (\citealt{Beckwith1990}) or from the wavelength at which the scattering polarisation peaks due to self-scattering\footnote{We would like to emphasize that the current model for self-scattering is based on perfectly spherical dust grains (\citealt{Kataoka2015}). This simplification of grain geometry, while being accurate for the description of scattering of the stellar light, is not applicable for description of scattering of the thermal dust emission. While the stellar light is intrinsically (almost perfectly) unpolarised, the thermal dust emission is intrinsically polarised. The intrinsic polarisation state of the emission depends on the grain geometry and, thus, changes the polarisation state after the scattering event (see Kirchschlager \& Bertrang, in prep.). Therefore, it is advisable to take  predictions of dust grain sizes based on current self-scattering models with caution.} (\citealt{Kataoka2015}), determining the characteristic wavelength offers a tool to derive the grain porosity in a protoplanetary disk. Then, surveying the wavelength dependence on the integrated intrinsic polarisation in the far-infrared and millimetre wavelength range provides  insight in the internal structure of the disk's building blocks. On the other hand, knowing the grain porosity from other techniques, e.g.~from the observation of the flipping of the polarisation direction by scattering at optical wavelengths (\citealt{Daniel1980, Kirchschlager2014}), or from the dust opacity index at \mbox{(sub-)}millimetre wavelengths (\citealt{Kataoka2014}), the intrinsic polarisation can be used to derive or verify the maximum grain size in the disk. 

\cite{Ohashi2018} detected polarisation fractions at wavelengths $\lambda=\unit[0.87]{mm}$ as high as $\unit[\sim15]{\%}$ in regions of the protoplanetary disk HD~142527. They interpreted this polarisation signal most likely due to grain alignment with magnetic fields  and thus as intrinsic polarisation. Since we find that intrinsic polarisation is decreasing with increasing porosity, the detection of \cite{Ohashi2018} sets an upper limit for the grain porosity in HD~142527. For a size distribution of silicate grains with maximum radii $a_\text{max}= \unit[100]{\mu m}$ and axis ratio $\nicefrac{c}{b}=1.5$ (Fig.~\ref{fig_pol_disk}), the intrinsic polarisations can take values up to $\unit[15]{\%}$ only for grains with porosities lower than $\mathcal{P}=0.7$. The maximum porosity is even lower for carbon or ice grains, for grains with lower axis ratio $\nicefrac{c}{b}$, or larger maximum grain sizes. For silicate grains with $\nicefrac{c}{b}=1.3$ and $a_\text{max}= \unit[100]{\mu m}$, only grain porosities up to $\mathcal{P}\le0.5$ are able to explain such high polarisation fractions, while the intrinsic polarisation from $\nicefrac{c}{b}=1.1$ spheroids is even too low for compact spheroids. Furthermore, if we assume that the grains are not perfectly aligned, the integrated polarisation signal is weaker pronounced, further decreasing the maximum porosity. 
The only possibility for the presence of larger grain porosities to be consistent with high polarisation fractions is to significantly increase the grain's axis ratio~$\nicefrac{c}{b}$. 

Observations of several other disks show polarisation fractions in the order of $\unit[\sim1]{\%}$ and  the most common explanations are self-scattering or intrinsic polarisation combined with depolarisation effects in optically thick regimes  (e.g.~\citealt{Kataoka2017, Girart2018, Sadavoy2018, Dent2019}). Consequently, the observed polarisation is not pure intrinsic polarisation in these disks, the determination of the characteristic wavelength $\lambda_\text{c}$ is not suitable to constrain grain porosity, and we are not able to completely rule out highly porous dust grains for these disks. 

Finally, we note that the change of the polarisation sign at certain wavelengths (Figs.~\ref{fig_pol_disk}  and \ref{fig_pol_disk_change1}) comes along with a drop of the polarisation, passes a null level and then rebuilds the polarised intensity, where the new orientation is flipped by 90 degree. This might explain the observed and still unexplained multi-wavelength pattern in HL~Tau, especially in waveband~3 (\citealt{Stephens2017}). Future observations of the linearly polarised gas emission will test this explanation. 

\section{Conclusions}
\label{401} 
We have calculated the intrinsic polarisation $P_\text{emi}$ for elongated porous dust grains in the far-infrared to millimetre wavelength range. Our results depend on the grain's effective radius $a_\text{eff}$ as the following:

\begin{enumerate}
 \item The amount of intrinsic polarisation $P_\text{emi}$ is decreasing with increasing porosity $\mathcal{P}$ for all particle shapes, an unexpected result as one would expect an increase of polarisation with increasing porosity. This result includes $\unit[100]{\mu m}$ grains, though they are impaired by the present oscillations in the corresponding wavelength range ($2\,a_\text{eff}\lesssim\lambda\lesssim10\,a_\text{eff}$).
 \item The intrinsic polarisation $P_\text{emi}$ of silicate grains is nearly independent of the wavelength $\lambda$ for $\lambda\gtrsim10\,a_\text{eff}$. For $2\,a_\text{eff}\lesssim\lambda\lesssim10\,a_\text{eff}$, oscillations occur for the absorption efficiency and the intrinsic polarisation. These oscillations are a result of enhanced interferences due to multiple reflection of the electromagnetic wave within the grain and enable even ``negative polarisation'' which implies a reversal of the polarisation direction.  
 For $\lambda\lesssim2\,a_\text{eff}$ (geometric optics regime), the emission becomes unpolarised. 
 \item  The intrinsic polarisation $P_\text{emi}$  increases with increasing aspherity of the particles, defined by the ratio of long to short semi-axis $\nicefrac{c}{b}$. The longer prolate grains or the flatter oblate grains, the larger the polarisation.
 \item Whether the grains are prolate or oblate has only a minor impact on the intrinsic polarisation $P_\text{emi}$. The amplitudes of the oscillations are slightly stronger for oblate grains compared to prolate grains but occur at similar wavelengths. Therefore, the ratio of long to short semi-axis $\nicefrac{c}{b}$ and the porosity $\mathcal{P}$ are the only parameters of the particle morphology that have significant influence on the polarisation $P_\text{emi}$.
 \item Besides grains made of astronomical silicate, the trends exist for carbon as well as icy grains though the wavelength dependence is slightly different.
\end{enumerate}
 
 We summarise for a grain size distribution in a protoplanetary disk:
 \begin{enumerate}
 \item Similar to the single grains, the intrinsic polarisation of the disk is decreasing with increasing porosity. $P_\text{emi,disk}$ is linearly increasing with $\log{(\lambda)}$ for $\lambda<\lambda_\text{c}$ and constant for larger wavelengths. 
 \item  The characteristic wavelength $\lambda_\text{c}$ is proportional to the maximum radius of the grain size distribution. For larger porosities, the   characteristic wavelength $\lambda_\text{c}$ is shifted to shorter wavelengths.
 \item  The  minimum grain size $a_\text{min}$ and the grain size exponent $\gamma$ have an impact on the steepness and the smoothness of the slope of the intrinsic polarisation for wavelengths  $\lambda<\lambda_\text{c}$, but not on the characteristic wavelength $\lambda_\text{c}$.
 \item  Surveying the wavelength dependence of the integrated intrinsic polarisation in the far-infrared and millimetre wavelength range provides a tool to derive the grain porosity or maximum grain size in a protoplanetary disk.
 \item Observed polarisation fractions in optically thin regions of protoplanetary disks (e.g.~for~HD~142527) set upper limits for the present grain porosities of up to $\unit[\sim70]{\%}$, rejecting highly porous (fluffy) grains unless the grain's axis ratio is  significantly larger than $1.5$. Intrinsic polarisation data of several grain shapes and dust materials are online available as supplementary material and allow the interpretation of future  polarisation observations.
 \end{enumerate}
 

\section*{Acknowledgements}
FK was supported by European Research Council Grant SNDUST ERC-2015-AdG-694520. GHMB and MF acknowledge funding from the European Research Council (ERC) under the European Union's Horizon 2020 research and innovation programme (grant agreement No. 757957).
  \appendix
  \setcounter{secnumdepth}{+2}
  \setcounter{section}{0} 
  \renewcommand\thesection{\Alph{section}}

\section{Absorption efficiency of porous dust grains of constant spatial extension}
\begin{figure*}
 \centering
 \includegraphics[trim=2.5cm 2.2cm 2cm 2.1cm,  clip=true,page=1,height=0.255\linewidth]{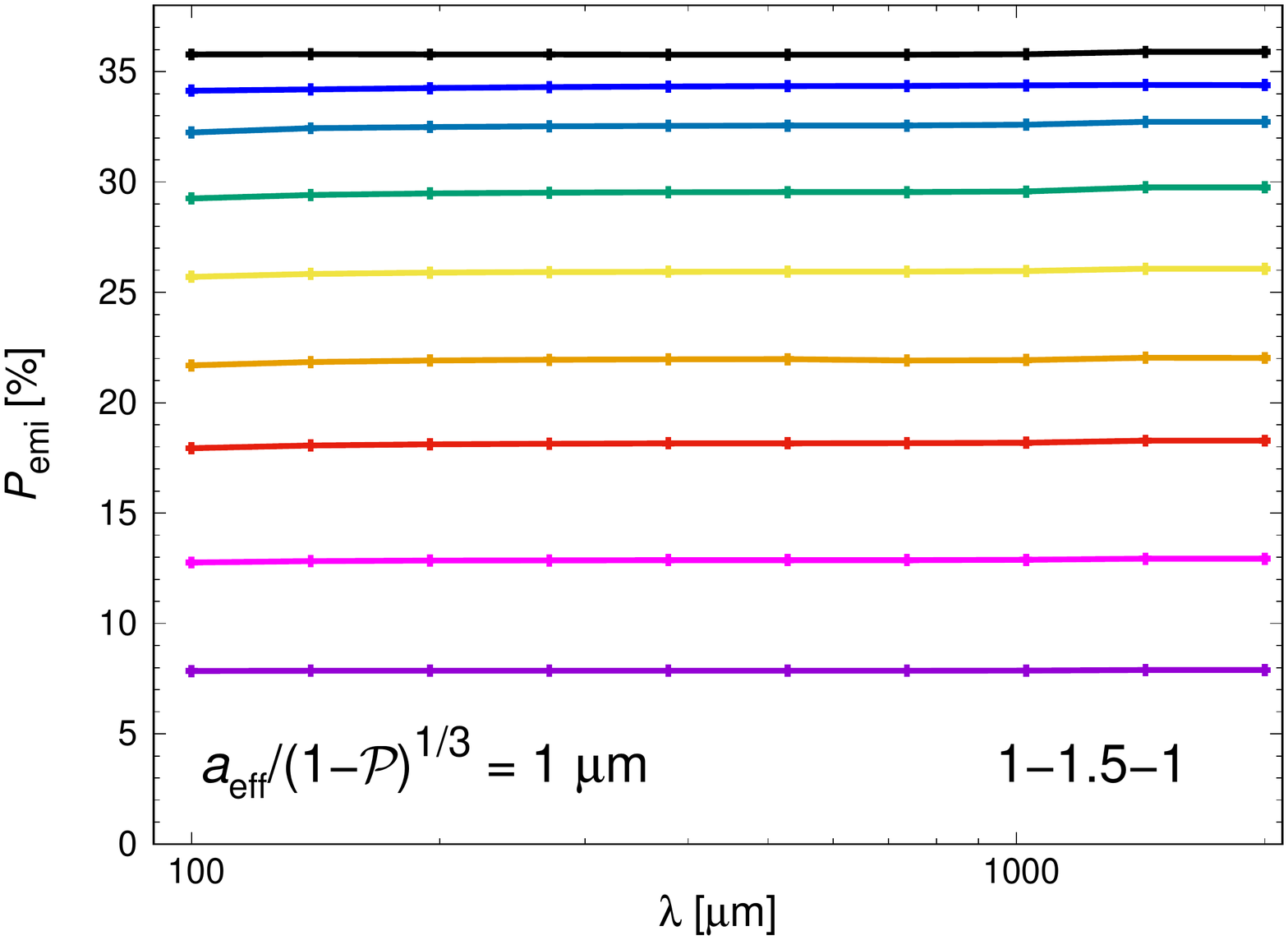}\hspace*{-0.01cm}
 \includegraphics[trim=3.8cm 2.2cm 2cm 2.1cm,  clip=true,page=2,height=0.255\linewidth]{Polarisation_prolate_oblate_same_shell.pdf}\hspace*{-0.01cm}
 \includegraphics[trim=3.8cm 2.2cm 2cm 2.1cm,  clip=true,page=3,height=0.255\linewidth]{Polarisation_prolate_oblate_same_shell.pdf}
\caption{Intrinsic polarisation for elongated porous silicates with fixed spatial extension, given by $a_\text{eff}/(1-\mathcal{P})^{1/3}$ (cf.~Fig.~\ref{fig_pol_emi}; \textit{top} row).}
\label{fig_compare_a_eff_and_a} 
\end{figure*}
\label{sec_same_shell} 

In Section~\ref{301} we compare grains with fixed effective radius $a_\text{eff}$ but different spatial extension. Here, we want to emphasize that the presented effects are not caused by comparing grains of different spatial extension. Therefore, we fix the radius of the encompassing spheroid to $a_\text{eff}/(1-\mathcal{P})^{1/3}=\unit[1]{\mu m},\unit[10]{\mu m}$, and $\unit[100]{\mu m}$, respectively, and determine the intrinsic polarisation $P_\text{emi}$ for prolate grains (1-1.5-1; astronomical silicate; Fig.~\ref{fig_compare_a_eff_and_a}). We can see a similar behaviour of $P_\text{emi}$ compared to the case of a fixed effective radius $a_\text{eff}$ (Fig.~\ref{fig_pol_emi}; \textit{top} row) though the oscillations are shifted by a factor of $(1-\mathcal{P})^{1/3}$ to shorter wavelengths.

\section{Comparison between T-Matrix Method (TMM) and Discrete Dipole Approximation (DDA)}
\label{A101}
\begin{figure*}
 \centering
 \includegraphics[trim=2.5cm 2.2cm 2cm 2.1cm,  clip=true,page=1,height=0.255\linewidth]{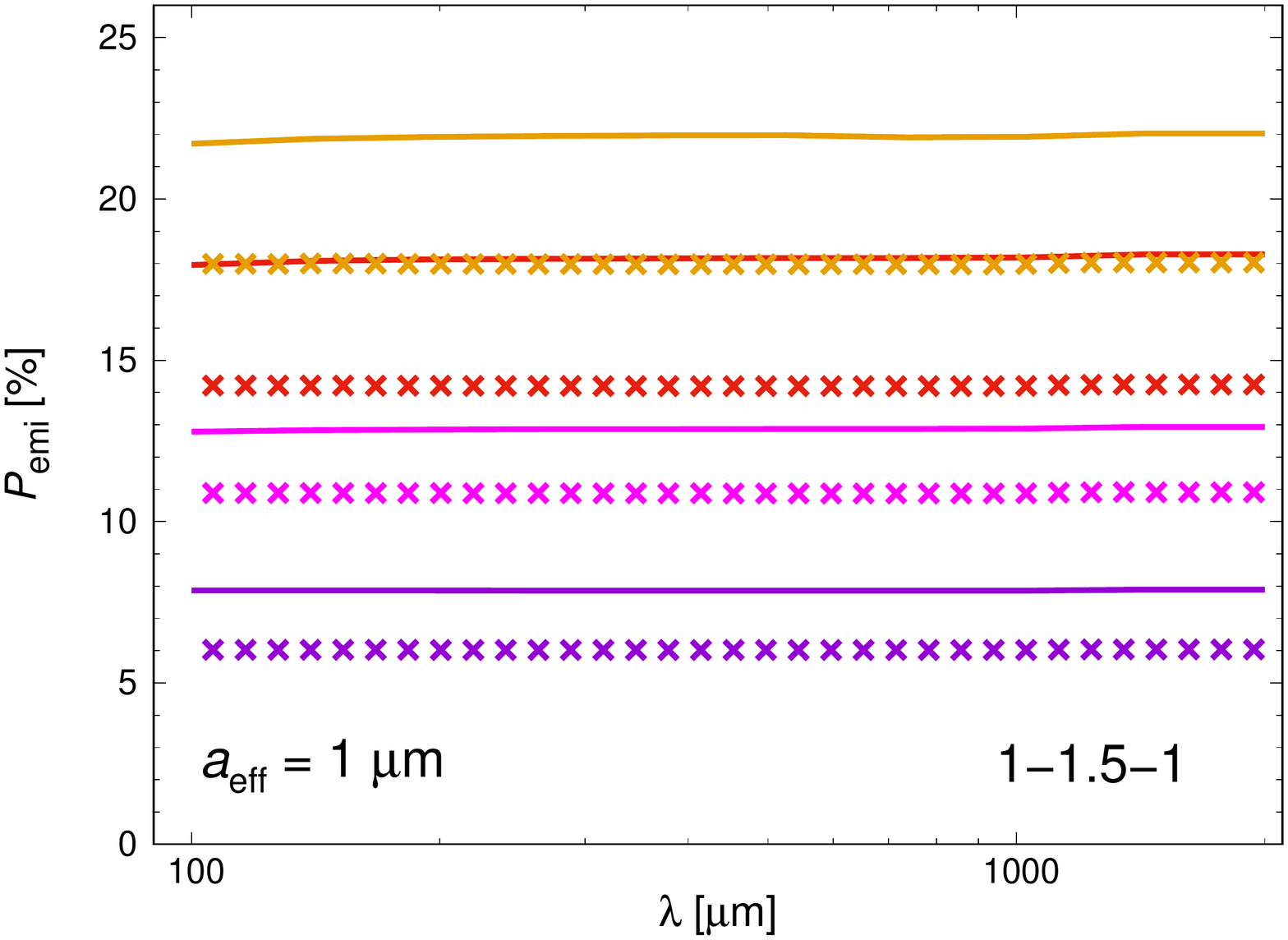}\hspace*{-0.01cm}
 \includegraphics[trim=3.8cm 2.2cm 2cm 2.1cm,  clip=true,page=2,height=0.255\linewidth]{Pol_emi_Sticks.pdf}\hspace*{-0.01cm}
 \includegraphics[trim=3.8cm 2.2cm 2cm 2.1cm,  clip=true,page=3,height=0.255\linewidth]{Pol_emi_Sticks.pdf}
\caption{Comparison of results obtained with DDA (solid lines) and TMM (crosses) for the intrinsic polarisation $P_\text{emi}$ as a function of wavelength $\lambda$. The silicate grains are prolate (1:1.5:1) and have effective dust radii of $a_\text{eff}=\unit[1]{\mu m}$, $\unit[10]{\mu m}$ and $\unit[100]{\mu m}$, respectively, and a porosity~$\mathcal{P}$ between $0.5$ and $0.8$.}
\label{fig_compare_DDSCAT_MST} 
\end{figure*}
 \begin{figure} 
  \centering
  \includegraphics[trim=4.8cm 0.8cm 5.4cm 2.3cm,  clip=true,page=1,height=0.55\linewidth]{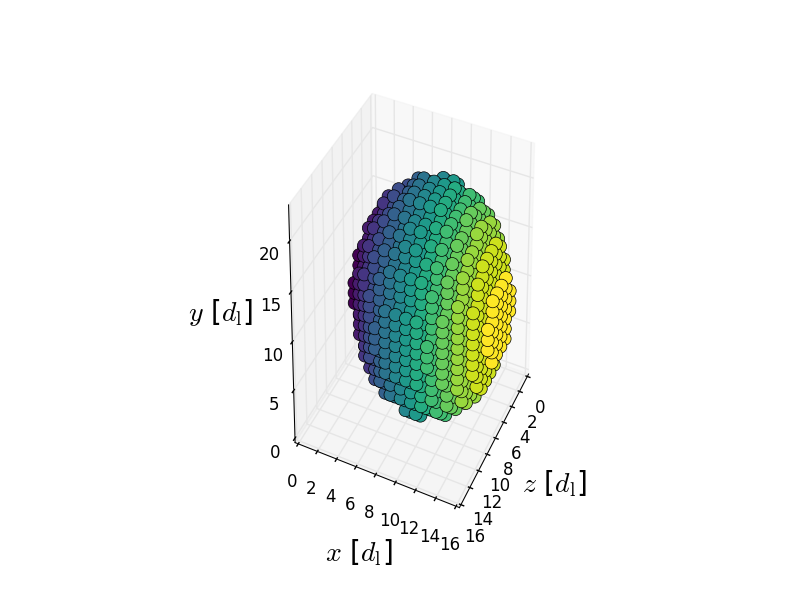}
  \includegraphics[trim=3.0cm -0.9cm 5.3cm 1.3cm,  clip=true,page=1,height=0.45\linewidth]{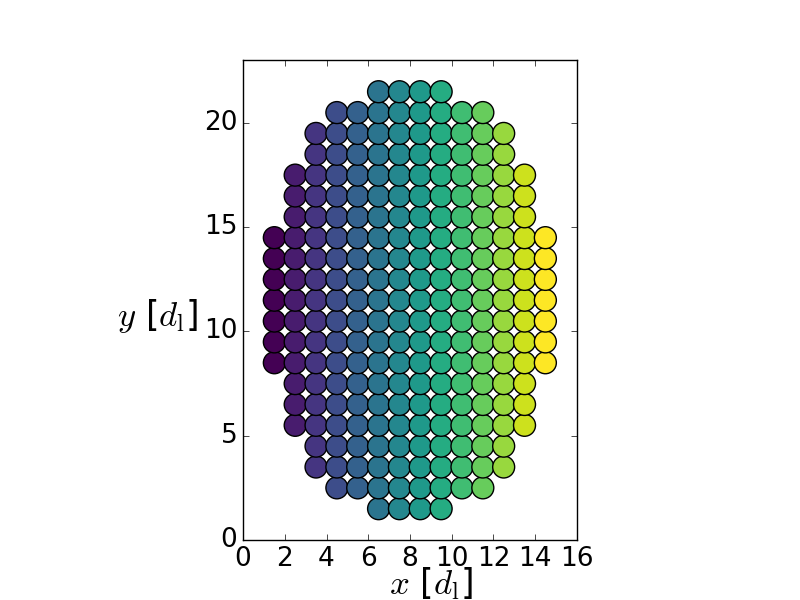}
 \caption{Morphology of an elongated porous dust grain (\mbox{1-1.5-1}) composed of 2204 equally sized spheres in a simple packing of spheres (\textit{left}). The inherent porosity (given by the voids between the spheres) is $\mathcal{P}=0.4764$. A single-layer of spherical monomers through the grain centre is shown \textit{right}.}
 \label{fig_morph_MST} 
 \end{figure}
In order to verify our results we consider a second calculation method that is unattached from calculations based on the DDA. We use the code  \textsc{MSTM}\footnote{\href{http://www.eng.auburn.edu/~dmckwski/scatcodes/}{http://www.eng.auburn.edu/$\sim$dmckwski/scatcodes/}} (Multiple Sphere \mbox{T-Matrix}; version 3.0) which applies the \mbox{T-Matrix} method (TMM) for an agglomeration of spherical monomers (\citealt{Mackowski1996}). Solving for the exact Mie solution of the single spheres, the optical properties of the agglomerate are determined on the superposition principle (\citealt{Mishchenko1996}). 

The three-dimensional particle shape of the elongated porous particle is replaced by a corresponding spatial distribution of $N_\text{s}$ spherical monomers. Because of the spherical geometry of the these monomers, stacking of them results in empty voids between the spheres which causes an inherent porosity of the agglomerate. For the sake of simplicity, we place $N_\text{s}$ equally sized spheres on a cubic grid (sc-lattice),\footnote{We note that even the close-packing of equal spheres (fcc- or hcp-lattice) has an inherent porosity of $\mathcal{P}\approx0.2595$. } forming an inherent porosity of $\mathcal{P}\approx0.4764$. Lower porosities are not considered.

We set $N_\text{s}=2204$ to form a prolate grain with shape \mbox{1-1.5-1} and porosity $\mathcal{P}\approx0.4764$ (Fig.~\ref{fig_morph_MST}). For larger numbers $N_\text{s}$, TMM~computations become time-consuming. We  then randomly remove 99 (520; 941; 1362) spherical monomers to generate a spheroidal grain with porosity $\mathcal{P}=0.5$ ($0.6$; $0.7$; $0.8$). These four particles are used to calculate the intrinsic polarisation $P_\text{emi}$ using TMM and to compare it with the corresponding results obtained with DDA (Fig.~\ref{fig_compare_DDSCAT_MST}). We note that the amounts of $P_\text{emi}$ derived from TMM calculations are  slightly smaller than that derived from DDA calculations for all grain sizes. However, considering the fact that the calculations have been conducted with totally different approximation methods, the presented results show a good agreement. Moreover, the simulations based on TMM confirm that the intrinsic polarisation of elongated dust grains is decreasing with increasing grain porosity.

  \bibliographystyle{mnras}
{\footnotesize
  \bibliography{Literature}

\begin{thebibliography}{}
\makeatletter
\relax
\def\mn@urlcharsother{\let\do\@makeother \do\$\do\&\do\#\do\^\do\_\do\%\do\~}
\def\mn@doi{\begingroup\mn@urlcharsother \@ifnextchar [ {\mn@doi@}
  {\mn@doi@[]}}
\def\mn@doi@[#1]#2{\def\@tempa{#1}\ifx\@tempa\@empty \href
  {http://dx.doi.org/#2} {doi:#2}\else \href {http://dx.doi.org/#2} {#1}\fi
  \endgroup}
\def\mn@eprint#1#2{\mn@eprint@#1:#2::\@nil}
\def\mn@eprint@arXiv#1{\href {http://arxiv.org/abs/#1} {{\tt arXiv:#1}}}
\def\mn@eprint@dblp#1{\href {http://dblp.uni-trier.de/rec/bibtex/#1.xml}
  {dblp:#1}}
\def\mn@eprint@#1:#2:#3:#4\@nil{\def\@tempa {#1}\def\@tempb {#2}\def\@tempc
  {#3}\ifx \@tempc \@empty \let \@tempc \@tempb \let \@tempb \@tempa \fi \ifx
  \@tempb \@empty \def\@tempb {arXiv}\fi \@ifundefined
  {mn@eprint@\@tempb}{\@tempb:\@tempc}{\expandafter \expandafter \csname
  mn@eprint@\@tempb\endcsname \expandafter{\@tempc}}}

\bibitem[\protect\citeauthoryear{{Beckwith}, {Sargent}, {Chini}  \&
  {Guesten}}{{Beckwith} et~al.}{1990}]{Beckwith1990}
{Beckwith} S.~V.~W.,  {Sargent} A.~I.,  {Chini} R.~S.,   {Guesten} R.,  1990,
  \mn@doi [Astronomical Journal] {10.1086/115385}, \href
  {http://adsabs.harvard.edu/abs/1990AJ.....99..924B} {99, 924}

\bibitem[\protect\citeauthoryear{{Bertrang} \& {Wolf}}{{Bertrang} \&
  {Wolf}}{2017}]{Bertrang2017b}
{Bertrang} G.~H.-M.,  {Wolf} S.,  2017, \mn@doi [\mnras]
  {10.1093/mnras/stx1066}, \href
  {http://adsabs.harvard.edu/abs/2017MNRAS.469.2869B} {469, 2869}

\bibitem[\protect\citeauthoryear{{Bertrang}, {Flock}  \& {Wolf}}{{Bertrang}
  et~al.}{2017}]{Bertrang2017a}
{Bertrang} G.~H.-M.,  {Flock} M.,   {Wolf} S.,  2017, \mn@doi [\mnras]
  {10.1093/mnrasl/slw181}, \href
  {http://adsabs.harvard.edu/abs/2017MNRAS.464L..61B} {464, L61}

\bibitem[\protect\citeauthoryear{{Blum} \& {Wurm}}{{Blum} \&
  {Wurm}}{2008}]{BlumWurm2008}
{Blum} J.,  {Wurm} G.,  2008, Annual Review of Astronomy {\&} Astrophysics, 46,
  21

\bibitem[\protect\citeauthoryear{{Cho} \& {Lazarian}}{{Cho} \&
  {Lazarian}}{2007}]{Cho2007}
{Cho} J.,  {Lazarian} A.,  2007, \mn@doi [Journal of Korean Astronomical
  Society] {10.5303/JKAS.2007.40.4.113}, \href
  {http://adsabs.harvard.edu/abs/2007JKAS...40..113C} {40, 113}

\bibitem[\protect\citeauthoryear{{Cox}, {Harris}, {Looney}, {Li}, {Yang},
  {Tobin}  \& {Stephens}}{{Cox} et~al.}{2018}]{Cox2018}
{Cox} E.~G.,  {Harris} R.~J.,  {Looney} L.~W.,  {Li} Z.-Y.,  {Yang} H.,
  {Tobin} J.~J.,   {Stephens} I.,  2018, \mn@doi [\apj]
  {10.3847/1538-4357/aaacd2}, \href
  {http://adsabs.harvard.edu/abs/2018ApJ...855...92C} {855, 92}

\bibitem[\protect\citeauthoryear{{Daniel}}{{Daniel}}{1980}]{Daniel1980}
{Daniel} J.-Y.,  1980, Astronomy \& Astrophysics, \href
  {http://adsabs.harvard.edu/abs/1980A%26A....87..204D} {87, 204}

\bibitem[\protect\citeauthoryear{{Dent}, {Pinte}, {Cortes}, {M{\'e}nard},
  {Hales}, {Fomalont}  \& {de Gregorio-Monsalvo}}{{Dent}
  et~al.}{2019}]{Dent2019}
{Dent} W.~R.~F.,  {Pinte} C.,  {Cortes} P.~C.,  {M{\'e}nard} F.,  {Hales} A.,
  {Fomalont} E.,   {de Gregorio-Monsalvo} I.,  2019, \mn@doi [\mnras]
  {10.1093/mnrasl/sly181}, \href
  {https://ui.adsabs.harvard.edu/abs/2019MNRAS.482L..29D} {482, L29}

\bibitem[\protect\citeauthoryear{{Dominik} \& {Tielens}}{{Dominik} \&
  {Tielens}}{1997}]{Dominik1997}
{Dominik} C.,  {Tielens} A.~G.~G.~M.,  1997, \mn@doi [The Astrophysical
  Journal] {10.1086/303996}, \href
  {http://adsabs.harvard.edu/abs/1997ApJ...480..647D} {480, 647}

\bibitem[\protect\citeauthoryear{{Draine}}{{Draine}}{2000}]{Draine2000}
{Draine} B.~T.,  2000, {The Discrete Dipole Approximation for Light Scattering
  by Irregular Targets}.
p.~131

\bibitem[\protect\citeauthoryear{{Draine}}{{Draine}}{2003a}]{Draine2003a}
{Draine} B.~T.,  2003a, The Astrophysical Journal, 598, 1017

\bibitem[\protect\citeauthoryear{{Draine}}{{Draine}}{2003b}]{Draine2003b}
{Draine} B.~T.,  2003b, The Astrophysical Journal, 598, 1026

\bibitem[\protect\citeauthoryear{{Draine} \& {Flatau}}{{Draine} \&
  {Flatau}}{1994}]{Draine1994}
{Draine} B.~T.,  {Flatau} P.~J.,  1994, Journal of the Optical Society of
  America A, 11, 1491

\bibitem[\protect\citeauthoryear{{Draine} \& {Flatau}}{{Draine} \&
  {Flatau}}{2010}]{Draine2010}
{Draine} B.~T.,  {Flatau} P.~J.,  2010, preprint

\bibitem[\protect\citeauthoryear{{Draine} \& {Goodman}}{{Draine} \&
  {Goodman}}{1993}]{DraineGood}
{Draine} B.~T.,  {Goodman} J.,  1993, The Astrophysical Journal, 405, 685

\bibitem[\protect\citeauthoryear{{Draine} \& {Weingartner}}{{Draine} \&
  {Weingartner}}{1996}]{Draine1996}
{Draine} B.~T.,  {Weingartner} J.~C.,  1996, The Astrophysical Journal, 470,
  551

\bibitem[\protect\citeauthoryear{{Draine} \& {Weingartner}}{{Draine} \&
  {Weingartner}}{1997}]{Draine1997}
{Draine} B.~T.,  {Weingartner} J.~C.,  1997, \mn@doi [\apj] {10.1086/304008},
  \href {http://adsabs.harvard.edu/abs/1997ApJ...480..633D} {480, 633}

\bibitem[\protect\citeauthoryear{{Fulle} et~al.,}{{Fulle}
  et~al.}{2015}]{Fulle2015}
{Fulle} M.,  et~al., 2015, \mn@doi [\apjl] {10.1088/2041-8205/802/1/L12}, \href
  {http://adsabs.harvard.edu/abs/2015ApJ...802L..12F} {802, L12}

\bibitem[\protect\citeauthoryear{{Girart} et~al.,}{{Girart}
  et~al.}{2018}]{Girart2018}
{Girart} J.~M.,  et~al., 2018, \mn@doi [\apjl] {10.3847/2041-8213/aab76b},
  \href {http://adsabs.harvard.edu/abs/2018ApJ...856L..27G} {856, L27}

\bibitem[\protect\citeauthoryear{{Gold}}{{Gold}}{1952}]{Gold1952}
{Gold} T.,  1952, \mn@doi [\mnras] {10.1093/mnras/112.2.215}, \href
  {http://adsabs.harvard.edu/abs/1952MNRAS.112..215G} {112, 215}

\bibitem[\protect\citeauthoryear{{Harris} et~al.,}{{Harris}
  et~al.}{2018}]{Harris2018}
{Harris} R.~J.,  et~al., 2018, \mn@doi [\apj] {10.3847/1538-4357/aac6ec}, \href
  {http://adsabs.harvard.edu/abs/2018ApJ...861...91H} {861, 91}

\bibitem[\protect\citeauthoryear{{Harrison} et~al.,}{{Harrison}
  et~al.}{2019}]{Harrison2019}
{Harrison} R.~E.,  et~al., 2019, arXiv e-prints, \href
  {http://adsabs.harvard.edu/abs/2019arXiv190506266H} {}

\bibitem[\protect\citeauthoryear{{Hull} et~al.,}{{Hull}
  et~al.}{2018}]{Hull2018}
{Hull} C.~L.~H.,  et~al., 2018, \mn@doi [\apj] {10.3847/1538-4357/aabfeb},
  \href {http://adsabs.harvard.edu/abs/2018ApJ...860...82H} {860, 82}

\bibitem[\protect\citeauthoryear{{J{\"a}ger}, {Mutschke}  \&
  {Henning}}{{J{\"a}ger} et~al.}{1998}]{Jaeger1998}
{J{\"a}ger} C.,  {Mutschke} H.,   {Henning} T.,  1998, \aap, \href
  {http://adsabs.harvard.edu/abs/1998A%26A...332..291J} {332, 291}

\bibitem[\protect\citeauthoryear{{Kataoka}, {Okuzumi}, {Tanaka}  \&
  {Nomura}}{{Kataoka} et~al.}{2014}]{Kataoka2014}
{Kataoka} A.,  {Okuzumi} S.,  {Tanaka} H.,   {Nomura} H.,  2014, \mn@doi
  [Astronomy \& Astrophysics] {10.1051/0004-6361/201323199}, \href
  {http://adsabs.harvard.edu/abs/2014A%26A...568A..42K} {568, A42}

\bibitem[\protect\citeauthoryear{{Kataoka} et~al.,}{{Kataoka}
  et~al.}{2015}]{Kataoka2015}
{Kataoka} A.,  et~al., 2015, \mn@doi [\apj] {10.1088/0004-637X/809/1/78}, \href
  {http://adsabs.harvard.edu/abs/2015ApJ...809...78K} {809, 78}

\bibitem[\protect\citeauthoryear{{Kataoka} et~al.,}{{Kataoka}
  et~al.}{2016}]{Kataoka2016}
{Kataoka} A.,  et~al., 2016, \mn@doi [\apjl] {10.3847/2041-8205/831/2/L12},
  \href {http://adsabs.harvard.edu/abs/2016ApJ...831L..12K} {831, L12}

\bibitem[\protect\citeauthoryear{{Kataoka}, {Tsukagoshi}, {Pohl}, {Muto},
  {Nagai}, {Stephens}, {Tomisaka}  \& {Momose}}{{Kataoka}
  et~al.}{2017}]{Kataoka2017}
{Kataoka} A.,  {Tsukagoshi} T.,  {Pohl} A.,  {Muto} T.,  {Nagai} H.,
  {Stephens} I.~W.,  {Tomisaka} K.,   {Momose} M.,  2017, \mn@doi [\apjl]
  {10.3847/2041-8213/aa7e33}, \href
  {http://adsabs.harvard.edu/abs/2017ApJ...844L...5K} {844, L5}

\bibitem[\protect\citeauthoryear{{Kirchschlager} \& {Wolf}}{{Kirchschlager} \&
  {Wolf}}{2013}]{Kirchschlager2013}
{Kirchschlager} F.,  {Wolf} S.,  2013, \mn@doi [Astronomy \& Astrophysics]
  {10.1051/0004-6361/201220486}, \href
  {http://adsabs.harvard.edu/abs/2013A%26A...552A..54K} {552, A54}

\bibitem[\protect\citeauthoryear{{Kirchschlager} \& {Wolf}}{{Kirchschlager} \&
  {Wolf}}{2014}]{Kirchschlager2014}
{Kirchschlager} F.,  {Wolf} S.,  2014, \mn@doi [Astronomy \& Astrophysics]
  {10.1051/0004-6361/201323176}, \href
  {http://adsabs.harvard.edu/abs/2014A%26A...568A.103K} {568, A103}

\bibitem[\protect\citeauthoryear{{Kothe}, {Blum}, {Weidling}  \&
  {G{\"u}ttler}}{{Kothe} et~al.}{2013}]{Kothe2013}
{Kothe} S.,  {Blum} J.,  {Weidling} R.,   {G{\"u}ttler} C.,  2013, \mn@doi
  [Icarus] {10.1016/j.icarus.2013.02.034}, \href
  {http://adsabs.harvard.edu/abs/2013Icar..225...75K} {225, 75}

\bibitem[\protect\citeauthoryear{{Langevin} et~al.,}{{Langevin}
  et~al.}{2016}]{Langevin2016}
{Langevin} Y.,  et~al., 2016, \mn@doi [\icarus] {10.1016/j.icarus.2016.01.027},
  \href {http://adsabs.harvard.edu/abs/2016Icar..271...76L} {271, 76}

\bibitem[\protect\citeauthoryear{{Lazarian} \& {Hoang}}{{Lazarian} \&
  {Hoang}}{2007}]{Lazarian2007}
{Lazarian} A.,  {Hoang} T.,  2007, \mn@doi [\mnras]
  {10.1111/j.1365-2966.2007.11817.x}, \href
  {http://adsabs.harvard.edu/abs/2007MNRAS.378..910L} {378, 910}

\bibitem[\protect\citeauthoryear{{Lee}, {Li}, {Ching}, {Lai}  \& {Yang}}{{Lee}
  et~al.}{2018}]{Lee2018}
{Lee} C.-F.,  {Li} Z.-Y.,  {Ching} T.-C.,  {Lai} S.-P.,   {Yang} H.,  2018,
  \mn@doi [\apj] {10.3847/1538-4357/aaa769}, \href
  {http://adsabs.harvard.edu/abs/2018ApJ...854...56L} {854, 56}

\bibitem[\protect\citeauthoryear{{Mackowski} \& {Mishchenko}}{{Mackowski} \&
  {Mishchenko}}{1996}]{Mackowski1996}
{Mackowski} D.~W.,  {Mishchenko} M.~I.,  1996, \mn@doi [Journal of the Optical
  Society of America A] {10.1364/JOSAA.13.002266}, \href
  {http://adsabs.harvard.edu/abs/1996JOSAA..13.2266M} {13, 2266}

\bibitem[\protect\citeauthoryear{{Mannel}, {Bentley}, {Schmied}, {Jeszenszky},
  {Levasseur-Regourd}, {Romstedt}  \& {Torkar}}{{Mannel}
  et~al.}{2016}]{Mannel2016}
{Mannel} T.,  {Bentley} M.~S.,  {Schmied} R.,  {Jeszenszky} H.,
  {Levasseur-Regourd} A.~C.,  {Romstedt} J.,   {Torkar} K.,  2016, \mn@doi
  [\mnras] {10.1093/mnras/stw2898}, \href
  {http://adsabs.harvard.edu/abs/2016MNRAS.462S.304M} {462, S304}

\bibitem[\protect\citeauthoryear{{Mathis}, {Rumpl}  \& {Nordsieck}}{{Mathis}
  et~al.}{1977}]{Mathis1977}
{Mathis} J.~S.,  {Rumpl} W.,   {Nordsieck} K.~H.,  1977, \mn@doi [The
  Astrophysical Journal] {10.1086/155591}, \href
  {http://adsabs.harvard.edu/abs/1977ApJ...217..425M} {217, 425}

\bibitem[\protect\citeauthoryear{{Milli} et~al.,}{{Milli}
  et~al.}{2019}]{Milli2019}
{Milli} J.,  et~al., 2019, arXiv e-prints, \href
  {https://ui.adsabs.harvard.edu/abs/2019arXiv190503603M} {}

\bibitem[\protect\citeauthoryear{{Mishchenko}, {Travis}  \&
  {Mackowski}}{{Mishchenko} et~al.}{1996}]{Mishchenko1996}
{Mishchenko} M.~I.,  {Travis} L.~D.,   {Mackowski} D.~W.,  1996, \mn@doi
  [\jqsrt] {10.1016/0022-4073(96)00002-7}, \href
  {http://adsabs.harvard.edu/abs/1996JQSRT..55..535M} {55, 535}

\bibitem[\protect\citeauthoryear{{Ohashi} et~al.,}{{Ohashi}
  et~al.}{2018}]{Ohashi2018}
{Ohashi} S.,  et~al., 2018, \mn@doi [\apj] {10.3847/1538-4357/aad632}, \href
  {http://adsabs.harvard.edu/abs/2018ApJ...864...81O} {864, 81}

\bibitem[\protect\citeauthoryear{{Ormel}, {Spaans}  \& {Tielens}}{{Ormel}
  et~al.}{2007}]{Ormel2007}
{Ormel} C.~W.,  {Spaans} M.,   {Tielens} A.~G.~G.~M.,  2007, \mn@doi [Astronomy
  \& Astrophysics] {10.1051/0004-6361:20065949}, \href
  {http://adsabs.harvard.edu/abs/2007A%26A...461..215O} {461, 215}

\bibitem[\protect\citeauthoryear{{Pinte} et~al.,}{{Pinte}
  et~al.}{2008}]{Pinte2008}
{Pinte} C.,  et~al., 2008, \mn@doi [\aap] {10.1051/0004-6361:200810121}, \href
  {https://ui.adsabs.harvard.edu/abs/2008A%26A...489..633P} {489, 633}

\bibitem[\protect\citeauthoryear{{Purcell} \& {Pennypacker}}{{Purcell} \&
  {Pennypacker}}{1973}]{PurcellPenny1973}
{Purcell} E.~M.,  {Pennypacker} C.~R.,  1973, \mn@doi [The Astrophysical
  Journal] {10.1086/152538}, \href
  {http://adsabs.harvard.edu/abs/1973ApJ...186..705P} {186, 705}

\bibitem[\protect\citeauthoryear{{Rao}, {Girart}, {Lai}  \& {Marrone}}{{Rao}
  et~al.}{2014}]{Rao2014}
{Rao} R.,  {Girart} J.~M.,  {Lai} S.-P.,   {Marrone} D.~P.,  2014, \mn@doi
  [\apjl] {10.1088/2041-8205/780/1/L6}, \href
  {http://adsabs.harvard.edu/abs/2014ApJ...780L...6R} {780, L6}

\bibitem[\protect\citeauthoryear{{Reinert}, {Mutschke}, {Krivov}, {L{\"o}hne}
  \& {Mohr}}{{Reinert} et~al.}{2015}]{Reinert2015}
{Reinert} C.,  {Mutschke} H.,  {Krivov} A.~V.,  {L{\"o}hne} T.,   {Mohr} P.,
  2015, \mn@doi [\aap] {10.1051/0004-6361/201424276}, \href
  {http://adsabs.harvard.edu/abs/2015A%26A...573A..29R} {573, A29}

\bibitem[\protect\citeauthoryear{{Sadavoy} et~al.,}{{Sadavoy}
  et~al.}{2018}]{Sadavoy2018}
{Sadavoy} S.~I.,  et~al., 2018, \mn@doi [\apj] {10.3847/1538-4357/aac21a},
  \href {http://adsabs.harvard.edu/abs/2018ApJ...859..165S} {859, 165}

\bibitem[\protect\citeauthoryear{{Segura-Cox}, {Looney}, {Stephens},
  {Fern{\'a}ndez-L{\'o}pez}, {Kwon}, {Tobin}, {Li}  \& {Crutcher}}{{Segura-Cox}
  et~al.}{2015}]{Seguracox2015}
{Segura-Cox} D.~M.,  {Looney} L.~W.,  {Stephens} I.~W.,
  {Fern{\'a}ndez-L{\'o}pez} M.,  {Kwon} W.,  {Tobin} J.~J.,  {Li} Z.-Y.,
  {Crutcher} R.,  2015, \mn@doi [\apjl] {10.1088/2041-8205/798/1/L2}, \href
  {http://adsabs.harvard.edu/abs/2015ApJ...798L...2S} {798, L2}

\bibitem[\protect\citeauthoryear{{Stephens} et~al.,}{{Stephens}
  et~al.}{2014}]{Stephens2014}
{Stephens} I.~W.,  et~al., 2014, \mn@doi [\nat] {10.1038/nature13850}, \href
  {http://adsabs.harvard.edu/abs/2014Natur.514..597S} {514, 597}

\bibitem[\protect\citeauthoryear{{Stephens} et~al.,}{{Stephens}
  et~al.}{2017}]{Stephens2017}
{Stephens} I.~W.,  et~al., 2017, \mn@doi [\apj] {10.3847/1538-4357/aa998b},
  \href {https://ui.adsabs.harvard.edu/abs/2017ApJ...851...55S} {851, 55}

\bibitem[\protect\citeauthoryear{{Takahashi}, {Machida}, {Tomisaka}, {Ho},
  {Fomalont}, {Nakanishi}  \& {Girart}}{{Takahashi}
  et~al.}{2019}]{Takahashi2019}
{Takahashi} S.,  {Machida} M.~N.,  {Tomisaka} K.,  {Ho} P.~T.~P.,  {Fomalont}
  E.~B.,  {Nakanishi} K.,   {Girart} J.~M.,  2019, \mn@doi [\apj]
  {10.3847/1538-4357/aaf6ed}, \href
  {http://adsabs.harvard.edu/abs/2019ApJ...872...70T} {872, 70}

\bibitem[\protect\citeauthoryear{{Tamura}, {Hough}, {Greaves}, {Morino},
  {Chrysostomou}, {Holland}  \& {Momose}}{{Tamura} et~al.}{1999}]{Tamura1999}
{Tamura} M.,  {Hough} J.~H.,  {Greaves} J.~S.,  {Morino} J.-I.,  {Chrysostomou}
  A.,  {Holland} W.~S.,   {Momose} M.,  1999, \mn@doi [The Astrophysical
  Journal] {10.1086/307953}, \href
  {http://adsabs.harvard.edu/abs/1999ApJ...525..832T} {525, 832}

\bibitem[\protect\citeauthoryear{{Tazaki} \& {Tanaka}}{{Tazaki} \&
  {Tanaka}}{2018}]{Tazaki2018}
{Tazaki} R.,  {Tanaka} H.,  2018, \mn@doi [\apj] {10.3847/1538-4357/aac32d},
  \href {http://adsabs.harvard.edu/abs/2018ApJ...860...79T} {860, 79}

\bibitem[\protect\citeauthoryear{{Tazaki}, {Lazarian}  \& {Nomura}}{{Tazaki}
  et~al.}{2017}]{Tazaki2017}
{Tazaki} R.,  {Lazarian} A.,   {Nomura} H.,  2017, \mn@doi [\apj]
  {10.3847/1538-4357/839/1/56}, \href
  {http://adsabs.harvard.edu/abs/2017ApJ...839...56T} {839, 56}

\bibitem[\protect\citeauthoryear{{Warren}}{{Warren}}{1984}]{Warren1984}
{Warren} S.~G.,  1984, \mn@doi [\ao] {10.1364/AO.23.001206}, \href
  {http://adsabs.harvard.edu/abs/1984ApOpt..23.1206W} {23, 1206}

\bibitem[\protect\citeauthoryear{{Ysard}, {Jones}, {Demyk}, {Bout{\'e}raon}  \&
  {Koehler}}{{Ysard} et~al.}{2018}]{Ysard2018}
{Ysard} N.,  {Jones} A.~P.,  {Demyk} K.,  {Bout{\'e}raon} T.,   {Koehler} M.,
  2018, \mn@doi [\aap] {10.1051/0004-6361/201833386}, \href
  {http://adsabs.harvard.edu/abs/2018A%26A...617A.124Y} {617, A124}

\bibitem[\protect\citeauthoryear{{van de Hulst}}{{van de
  Hulst}}{1981}]{vandeHulst1981}
{van de Hulst} H.~C.,  1981, {Light scattering by small particles}

\makeatother
\end{thebibliography}
}
\label{lastpage}

\end{document}